\documentclass[11pt,a4paper]{article}
\usepackage{jcappub}
\usepackage{slashed}
\usepackage{braket}
\usepackage{graphicx}
\usepackage{hyperref}
\usepackage{float}
\usepackage{amssymb}
\usepackage{mathrsfs}
\usepackage{afterpage}

\usepackage[normalem]{ulem}

\newcommand{\amd}{a_{\rm md}}
\newcommand{\anu}{a_\nu}

\newcommand{\Ci}{{\mathrm{Ci}}}
\newcommand{\cL}{c_{\rm L}}
\newcommand{\classcode}{{\sc class}}
\newcommand{\dcb}{\delta_{\rm cb}}

\newcommand{\dfdlnp}{\frac{d \bar f}{d \log p}}
\newcommand{\dfmu}{\delta \! f}

\newcommand{\dm}{\delta_{\rm m}}
\newcommand{\dnu}{\delta_\nu}

\newcommand{\Ec}{{\mathcal E}}
\newcommand{\ehdm}{{\textrm{\sc{ehdm}}}}
\newcommand{\fastnuf}{{\sc fast}-$\nu f$}
\newcommand{\Fc}{{\mathcal F}}
\newcommand{\fcb}{f_{\rm cb}}
\newcommand{\FE}{{\mathcal F}_{\rm E}}
\newcommand{\FL}{{\mathcal F}_{\rm L}}
\newcommand{\FNLO}{{\mathcal F}_{\rm NLO}}
\newcommand{\Fnr}{{\mathcal F}_{\rm nr}}

\newcommand{\fnu}{f_\nu}
\newcommand{\gammaEM}{\gamma_{\textrm{\sc{em}}}}
\newcommand{\Hc}{{\mathcal H}}
\newcommand{\Hco}{{\mathcal H}_0}
\newcommand{\hdm}{{\textrm{\sc{hdm}}}}
\newcommand{\IEc}{{\mathcal I}^{({\rm E})}}
\newcommand{\ILc}{{\mathcal I}^{({\rm L})}}
\newcommand{\kfs}{k_{\rm fs}}

\newcommand{\knr}{k_{\rm nr}}
\newcommand{\Mnu}{M_\nu}

\newcommand{\ns}{n_{\rm s}}

\newcommand{\Obo}{\Omega_{\mathrm{b},0}}

\newcommand{\Od}{\Omega_\mathrm{d}}

\newcommand{\Odo}{\Omega_{\mathrm{d},0}}
\newcommand{\Of}{\Omega_f}

\newcommand{\Ofo}{\Omega_{f,0}}

\newcommand{\OKo}{\Omega_{K,0}}
\newcommand{\Om}{\Omega_\mathrm{m}}

\newcommand{\Omo}{\Omega_{\mathrm{m},0}}

\newcommand{\Ono}{\Omega_{\nu,0}}

\newcommand{\PLeg}{{\mathcal P}}
\newcommand{\psieds}{\psi_{\rm EdS}}
\newcommand{\RE}{{\mathcal R}_{\rm E}}
\newcommand{\Reds}{R_{\rm EdS}}

\newcommand{\RL}{{\mathcal R}_{\rm L}}

\newcommand{\SEEdS}{SE-EdS}
\newcommand{\Si}{{\mathrm{Si}}}
\newcommand{\si}{{\mathrm{si}}}
\newcommand{\sinc}{{\mathrm{sinc}}}
\newcommand{\sinit}{s_{\rm in}}
\newcommand{\tconf}{{\mathcal T}}

\newcommand{\Tnuo}{T_{\nu,0}}
\newcommand{\UEc}{{\mathcal U}^{({\rm E})}}
\newcommand{\ULc}{{\mathcal U}^{({\rm L})}}
\newcommand{\VEc}{{\mathcal V}^{({\rm E})}}
\newcommand{\VLc}{{\mathcal V}^{({\rm L})}}
\newcommand{\wLJ}{w_{\rm LJ}}
\newcommand{\woai}{|w_0|^{-1}}

\newcommand{\XLJ}{X_{\rm LJ}}
\newcommand{\yin}{y_{\rm in}}

\newcommand{\EQ}[1]{Eq.~(\ref{#1})}
\newcommand{\EQEQ}[2]{Eqs.~(\ref{#1},~\ref{#2})}
\newcommand{\EQEQEQ}[3]{Eqs.~(\ref{#1},~\ref{#2},~\ref{#3})}

\title{Planting neutrino \SEEdS{}: A ${\mathbf{<3\%}}$ accurate clustering approximation for neutrinos and other hot dark matter}

\author[a,b,c]{Amol Upadhye}
\author[c]{and Yvonne Y.~Y.~Wong}
\affiliation[a]{South-Western Institute for Astronomy Research,
  Yunnan University,
  Kunming, Yunnan 650500, People's Republic of China}
\affiliation[b]{Yunnan Key Laboratory of Survey Science,
  Yunnan University,
  Kunming, Yunnan 650500, People's Republic of China}
\affiliation[c]{Sydney Consortium for Particle Physics and Cosmology,
  School of Physics, The University
  of New South Wales, Sydney NSW 2052, Australia}

\emailAdd{a.upadhye@ynu.edu.cn}
\emailAdd{yvonne.y.wong@unsw.edu.au}

\abstract{
The age of the cosmic neutrino is upon us.  Aside from their fundamental importance to particle physics and the recent tensions in their mass scale $\Mnu = \sum m_\nu$, neutrinos' degeneracies with other cosmological parameters in analyses of scale-dependent cosmic clustering will significantly affect the estimates of these parameters,  making accurate, inexpensive, easily-implemented approximations essential tools for theoretical cosmology.  
We present a new linear approximation, SE-EdS, applicable to non-relativistic neutrinos and other hot dark matter, which is exact in the linear Einstein-de Sitter case, and $< 3\%$ accurate for a broad range of flat and non-flat cosmologies characterized by parameters $0.2 \leq \Omega_{{\rm m},0} \leq 1$, $0 \leq \Omega_{\nu,0} \leq 0.018$, $-0.1 \leq \Omega_{K,0} \leq 0.1$, and $w_0 \leq -0.6$, at all length scales.  This broad coverage in $\Om$ makes SE-EdS ideal for N-body simulations, whose redshift range $z \lesssim 100$ demands accuracy for time-dependent $\Omega_{\rm m}$ between unity and the preferred late-time values $\sim 0.3$.  SE-EdS may be used to paint neutrinos into simulations in Fourier space, where its $<3\%$ error and neutrinos' $\leq 3\%$ matter fraction guarantees $<0.1\%$ errors in the total matter density contrast, thereby including neutrino effects upon clustering at negligible computational cost.  We quantify the approximation's accuracy; extend it to a broad range of cosmologies; and explore its correspondence with higher-order corrections to existing linear response schemes.  A simple implementation is available online at {\tt https://codeberg.org/upadhye/SE-EdS}~.
}

\begin{document}

\begin{flushright}
	{\large \tt CPPC-2026-07}
\end{flushright}	

\maketitle

\section{Introduction}
\label{sec:int}

Beneath its calm, cool exterior, the standard cosmological model trembles.  As recent data~\cite{Planck:2018vyg,DES:2024jxu,DESI:2024hhd,DESI:2025ejh,SPT-3G:2025bzu} have tightened constraints on its basic parameters, such as the density fractions of cold dark matter (CDM) and the cosmological constant, tensions have emerged.  One is between different constraints on the absolute neutrino mass scale $\Mnu$.  Although laboratory oscillation experiments~\cite{DayaBay:2022orm,Esteban:2024eli,JUNO:2025gmd} cannot determine $\Mnu$ directly, the sum of their long-standing, extremely well-measured mass splittings implies a lower bound $\Mnu \geq 59$~meV.  This is above the cosmological upper bound of Ref.~\cite{DESI:2025ejh}, $53$~meV, though this latter value assumes a cosmological constant dark energy, and there may be internal inconsistencies among the individual cosmological probes comprising it~\cite{Ong:2026tta,Karim:2026jcx}.  A second tension, persisting for over a decade, is between sets of experiments preferring low ($\approx 67$~km/sec/Mpc) and high ($\approx 74$~km/sec/Mpc) values for the Hubble parameter $\Hco$~\cite{Cai:2026swf,Efstathiou:2026fax,Li:2026asg,Schoneberg:2026vaf}.  A recent survey~\cite{Freedman:2024eph} searching for agreement among three different $\Hco$ measurements, instead found one consistent with the low values, another consistent with the high ones, and yet another in the middle.  A third tension is between the amplitude $\sigma_8$ of linear clustering preferred by early-universe probes such as the cosmic microwave background (CMB) and that preferred at late times~\cite{Leauthaud:2016jdb,Zu:2025lrk,Pantos:2026koc}.  

Since neutrinos suppress small-scale, late-time structure formation, and since non-standard numbers of neutrinos in the early universe have been proposed as solutions to the Hubble tension~\cite{DiValentino:2021izs}, neutrinos are central to the first tension and potentially related to the other two.  Moreover, neutrinos are degenerate with other cosmological parameters, and the strength of modern constraints means that they cannot be excluded from consideration.  In fact, degeneracies betweeen neutrinos and the total matter density fraction, the dark energy equation of state, and the optical depth to reionization are integral to several proposed solutions of the $\Mnu$ tension~\cite{Kamionkowski:2022pkx,Jhaveri:2025neg,Sailer:2025lxj}.  Thus, neutrinos must be included in analyses of scale-dependent cosmological clustering even when the focus is on other parameters.

However, rigorously including neutrinos in non-linear clustering calculations is theoretically daunting and numerically expensive, whether the method is an N-body particle simulation, a non-linear perturbation theory, or an effective field theory~\cite{Brandbyge:2009ce,Banerjee:2018bxy,Elbers:2020lbn,Chen:2022cgw,Upadhye:2024ypg,Garny:2022fsh}.  The chief difficulty is the neutrinos' large thermal velocity dispersion, which forces us to track their full six-dimensional phase space, as opposed to merely the three-dimensional spatial distribution of cold matter.  For example, the root-mean-squared speed of a cosmologically-typical $50$~meV neutrino is $\approx 1 \%$ of the speed of light today, and considerably higher at the highest redshifts accessible to galaxy and lensing surveys.  Such fast-moving neutrinos exceed the escape velocities of galaxies and small galaxy clusters, while being captured by the largest clusters, leading to a scale-dependent growth suppression essential to the phenomenology of cosmological neutrinos.  Thus, tracking neutrinos can easily consume the lion's share of computational resources in a simulation campaign originally intended for the study of some different effect.  This creates an opening for approximate methods that can capture the neutrino clustering behavior in inexpensive ways.

Into this breach has stepped a host of linear response methods.  Broadly speaking, linear response couples linear perturbation theory for neutrinos to a non-linear treatment (e.g., an N-body simulation) of the CDM and baryonic fluids~\cite{Brandbyge:2008js,Ali-Haimoud:2012fzp}. A linear-response solution is sought for the neutrino density fluctuations given the non-linear CDM and baryon perturbations, which is then ``painted'' into the N-body simulations at each time step to evolve the CDM and baryon components further.
As reported in the existing literature, these methods have varying degrees of speed, accuracy, and ease of implementation in a N-body simulation.  A new FFT-based approach for neutrinos which may be relativistic at times of interest is described in Refs.~\cite{Ji:2022iji,Lee:2025zym,Lee:2025vgv}.  For fully non-relativistic neutrinos, the quickest and easiest approximation is the Super-Easy (SE) linear response of Refs.~\cite{Ringwald:2004np,Wong:2008ws,Chen:2020kxi}, which uses a simple rational function to interpolate the neutrino-to-matter density contrast ratio $\dnu/\dm$ between leading-order expansions in the large-scale clustering limit (i.e., $k/k_{\rm fs} \ll 1$, where $k_{\rm fs}$ is the free-streaming wavenumber) and the small-scale free-streaming limit (i.e., $k/k_{\rm fs} \gg 1)$. 
Its two flaws are that it is only $\gtrsim 10\%$ accurate in the interpolation region between these limits, with even larger errors at redshifts of a few, and that it requires a sound speed which formally diverges for the Bose-Einstein distribution, making this method applicable to neutrinos but not a general hot dark matter (HDM).  Although the latter flaw has been fixed by the Generalized Super Easy approach of Ref.~\cite{Pierobon:2024kpw}, which interpolates the limiting solutions one single momentum at a time, the former flaw is more difficult to solve, in part due to non-perturbative higher-order terms in the clustering-limit expansion.  

We take a new approach in this Article.  Rather than interpolating approximate, limiting solutions of $\dnu/\dm$, we form a ratio of $\dnu/\dm$ with the exact Einstein-de Sitter (EdS) solution in the neutrino test-particle limit and develop a fitting function for this ratio in terms of the growth function.  Since the EdS solution is exact, non-perturbative terms in the small $k/k_{\rm fs}$ expansion are automatically accounted for to all orders.  Further, the EdS $\dnu/\dm$ already agrees with $\dnu/\dm$ for currently-preferred cosmologies at the $10\%-20\%$ level at all scales, reducing the burden placed upon the fitting function.  We find a pair of SE-like rational function fits for the ratio of $\dnu/\dm$ for a single neutrino thermal momentum to its EdS equivalent.  As they arise from the application of SE to EdS, we call these the \SEEdS{} fits.  Remarkably simple, they describe the deviation of $\dnu/\dm$ from its EdS counterpart in terms of the function $1 - D(a)/a$, with $D(a)$ the large-scale growth factor.  A natural advantage of this approach is that it pushes $\dnu/\dm$ to its EdS equivalent at early times, $a \lesssim 0.1$, when the universe is typically matter-dominated, making \SEEdS{} more accurate at early times.

Our study of the \SEEdS{} fits begins with the case of spatially-flat cosmologies with constant dark energy equations of state, for which an exact solution for $D(a)$ is known.   Thus our fits can be written down in closed form.  Tuning one parameter at a time away from a standard fiducial cosmology, we find accuracy to $<3\%$ for a broad range of parameters.  Moreover, since the \SEEdS{} fits do not depend directly upon the cosmological parameters, but only $D(a)$, they are very easy to generalize to arbitrary dark energy equations of state and other models.  We thus extend our tests to spatial curvature and to evolving dark energy equations of state, for which $D(a)$ may be computed numerically, characterizing their accuracy for a broad range of parameters.  Considering a phenomenological approximation to models with couplings between CDM and dark energy, we confirm the accuracy of \SEEdS{} to $<3\%$ for a wide range of modifications to the matter growth rate.  We then proceed to test \SEEdS{} as a linear response method, showing it to be much more accurate than SE fits in the linear regime $k \lesssim 0.1~h/$Mpc, and comparably accurate in the non-linear regime.  

Finally, we compare \SEEdS{} fits to an alternative approach based upon higher-order corrections to the free-streaming limit of $\dnu/\dm$.  For the $\Lambda$CDM model at early times $a \sim 0.1$, when the matter density fraction $\Om(a)$ is close to unity, the two agree remarkably well.  However, \SEEdS{} is more accurate at late times, with $\Om(a) \approx 0.3$, and much more accurate for very negative dark energy equations of state.  This latter improvement is due to the perturbative expansion parameter ceasing to be small for large negative equations of state.  Thus \SEEdS{} is effectively a regularization of the next-to-leading-order correction which improves its accuracy and extends it to a wider range of dark energy models.

This Article is organized as follows.  After summarizing known results in Sec.~\ref{sec:bkg}, we discuss exact EdS solutions in the EdS-test-HDM case in Sec.~\ref{sec:eds}.  Section~\ref{sec:fit} introduces and tests our \SEEdS{} fits after discussing three different limits in which they approach the exact EdS solution.  Readers primarily interested in painting neutrinos into their N-body simulations may skip directly to Sec.~\ref{sub:fit:paint}, which we have written to be self-contained. In Sec.~\ref{sec:gen}, we test the fits for a broader range of cosmologies, and we explore their utility for linear response.  Section~\ref{sec:nlo} explores the correspondence between \SEEdS{} and higher-order free-streaming corrections, after which Sec.~\ref{sec:con} concludes.

\section{Background}
\label{sec:bkg}

In order to keep the Article self-contained, and to specify our notational conventions, this Section collects several standard results in homogeneous cosmology and linear perturbation theory.  Section~\ref{sub:bkg:homog} provides exact formulae for the homogeneous evolution and linear growth in a model containing only matter and a dark energy with a constant equation of state, then shows how to generalize beyond this case.  Linear perturbation theory for the neutrino is outlined in Sec.~\ref{sub:bkg:nupt}, in the Eulerian (fixed-coordinate) and Lagrangian-momentum (fixed initial momentum) cases, as well as an approximate fit.  The effective distribution function, allowing multiple neutrinos and other HDM species to be represented by a single effective species, is summarized in Sec.~\ref{sub:bkg:ehdm}.

Eulerian and Lagrangian fluid dynamics are two equivalent descriptions of the same physics.  Since the approach of Refs.~\cite{Dupuy:2013jaa,Dupuy:2014vea} used here is Lagrangian only in momentum-space, integrating the Eulerian and Lagrangian-momentum neutrino density contrasts over momenta will result in the same total neutrino density contrast.  At a technical level, each has its own advantages for our work, as will become apparent in subsequent sections.  Eulerian perturbation theory will have a more accurate next-to-leading-order correction which will provide useful insight into our later fitting functions.  Lagrangian-momentum perturbation theory has a somewhat simpler density contrast, and it has been expanded into a non-linear perturbation theory for HDM in Refs.~\cite{Chen:2022cgw,Upadhye:2024ypg}.  We discuss both, here and in later sections.

\subsection{Homogeneous universe}
\label{sub:bkg:homog}

We are concerned here with the behavior of HDM at late times, $z \lesssim 100$.  For most of this Article, we restrict ourselves to spatially-flat ``$w_0$CDM'' cosmologies containing only: non-relativistic matter, of density fraction $\Omo$ today; and a single dark energy species of density fraction $\Odo=1-\Omo$ and constant equation of state $w_0 = \bar P_{\rm d} / \bar\rho_{\rm d}$, with $\bar P_{\rm d}$ and $\bar\rho_{\rm d}$, respectively, the dark energy pressure and energy density.  At the end of this subsection, we briefly discuss more general cosmological models, including spatial curvature as well as time-dependent dark energy equations of state.


First, consider $w_0$CDM models.  Exact solutions for the time variables and the large-scale growth factor are available for these models, and they are general enough that we can explore the impact of dark energy upon our HDM predictions.

The time-dependent matter and dark energy density fractions in $w_0$CDM are:
\begin{equation}
  \Om(a) = \frac{1}{1+a^{-3w_0}/r_0}, \quad
  \Od(a) = \frac{a^{-3w_0}/r_0}{1+a^{-3w_0}/r_0}, \quad
  \mathrm{where}\quad r_0 := \frac{\Omo}{\Odo}=\frac{\Omo}{1-\Omo}.
  \label{e:Oma_Oda_w0cdm}
\end{equation}
The conformal time $\tconf$ and superconformal time $s$, defined by $ds = d\tconf/a$, are given by
\begin{eqnarray}
  \tconf(a)
  &=&
  \frac{2\Om(a)^{\tfrac{1}{6w_0}}}{\sqrt{\Om(a)\Hc(a)^2}}
  ~{}_2F_1\left[ -\frac{1}{6w_0}, \frac{1}{2} -\frac{1}{6w_0},
    1-\frac{1}{6w_0}, \Od(a)\right],
  \label{e:tconf_w0cdm}
  \\
  s(a)
  &=&
  -\frac{2\Om(a)^{\tfrac{1}{6w_0}}}{a\sqrt{\Om(a)\Hc(a)^2}}
  ~{}_2F_1\left[ \frac{1}{6w_0}, \frac{1}{2} + \frac{1}{6w_0},
    1 + \frac{1}{6w_0}, \Od(a) \right].
  \label{e:s_w0cdm}
\end{eqnarray}
The growth factor $D(a)$ and its derivative are:
\begin{eqnarray}
  G(a)
  &:=& \frac{D(a)}{a} =
  \Om(a)^{-\frac{1}{3w_0}}
  ~{}_2F_1\left[ -\frac{1}{3w_0}, \frac{1}{2} - \frac{1}{3w_0},
    1 - \frac{5}{6w_0}, \Od(a) \right],
  \label{e:G_w0cdm}
  \\
  \frac{dD}{da}
  &=&
  \Om(a)^{1-\frac{1}{3w_0}}
  ~{}_2F_1\left[ -\frac{1}{3w_0}, \frac{1}{2} - \frac{1}{3w_0},
    1 - \frac{5}{6w_0}, \Od(a) \right]
  \ldots\nonumber\\
  &~&
  + \frac{2-3w_0}{5-6w_0}\Od(a) \Om(a)^{1-\frac{1}{3w_0}}
  ~{}_2F_1\left[1 -\frac{1}{3w_0}, \frac{3}{2} - \frac{1}{3w_0},
    2 - \frac{5}{6w_0}, \Od(a) \right].
  \label{e:dDda_w0cdm}
\end{eqnarray}
The general form of the linear density contrast was derived in Refs.~\cite{Lee:2009if,Lee:2009gb,Lee:2009ft}, while the growing mode solutions above, as well as the preceding time variables, are from Ref.~\cite{Upadhye:2025glk}.  In models where HDM makes the growth scale-dependent, we use $D(a)$ and $G(a)$ to represent the large-scale limits given above, in which HDM behaves like CDM.  In the EdS limit, $r_0\rightarrow \infty$, $\Od(a)$ vanishes, all of the above hypergeometric functions reduce to unity, and we recover the familiar results $\tconf_{\rm EdS} = 2\sqrt{a}/\Hco$, $s_{\rm EdS} = -2/(\Hco \sqrt{a})$, $G_{\rm EdS}=1$, and $dD_{\rm EdS}/da=1$.


Next, consider a more general cosmology, with multiple fluids and time-varying equations of state.  The conformal Hubble parameter in such a universe may be written
\begin{eqnarray}
  \frac{\Hc^2}{\Hco^2}
  &=&
  \sum_{{\rm fluids}~f} \Ofo \Ec_f(a),
  \qquad
  \frac{d \log\Hc}{d \log a}
  = \frac{\Hco^2}{2\Hc^2} \sum_{{\rm fluids}~f} \Ofo \frac{d\Ec_f(a)}{d\log a}
  \\
  \Ec_f(a)
  &=&
  \frac{a^2 \bar\rho_f(a)}{\bar\rho_{f,0}}
  = \left\{
  \begin{array}{ll}
    a^{-1-3w_0} & \mathrm{for}~w(a)=w_0 \\
    a^{-1-3w_0-3w_a}e^{-3w_a(1-a)} & \mathrm{for}~w(a) = w_0 + w_a(1-a) \\
    a^{-1} \sqrt{\tfrac{1-v(1)^2}{1-v(a)^2}} & \mathrm{uniform~speed}~v(a) 
  \end{array}
  \right.
\end{eqnarray}
The time-dependent density fraction of fluid $f$ is then $\Of(a) = \Ofo \Hco^2 \Ec_f(a) / \Hc(a)^2$.  Thus, for example, the matter density is $\Om(a) = \Omo \Hco^2 / (a \Hc^2)$, implying $\Om(a) \Hc(a)^2 = \Omo \Hco^2/a$.  This may be used to simplify \EQEQ{e:tconf_w0cdm}{e:s_w0cdm} further.

We use the approach of Linder and Jenkins, Ref.~\cite{Linder:2003dr}, which treats every component other than matter as a non-clustering perfect fluid characterized by an equation of state $w(a)$; thus, curvature and radiation have constant equations of state $w=-1/3$ and $w=1/3$, respectively.  Then the equation governing the matter growth factor depends upon two functions characterizing these non-matter fluids:
\begin{equation}
  \wLJ(a)
  :=
  -\frac{1/3}{1-\Om(a)} \left( \frac{d\log\Hc^2}{d\log a} + 1 \right),
  \quad
  \XLJ(a)
  :=
  \frac{\Om(a)}{1-\Om(a)}.
  \label{e:wLJ_XLJ}
\end{equation}
The resulting growth factor equation of Ref.~\cite{Linder:2003dr} is
\begin{equation}
  \frac{d^2 G}{da^2}
  + \left[ \frac{7}{2} - \frac{3}{2}\frac{\wLJ(a)}{1+\XLJ(a)} \right]
  \frac{1}{a} \frac{d G}{da}
  + \left[ \frac{3}{2} \frac{1-\wLJ(a)}{1+\XLJ(a)} \right] \frac{G}{a^2}
  = 0
  \label{e:GLJ}
\end{equation}
As above, this $G(a)$ applies to large scales, on which  HDM clusters like CDM.  Integrating numerically for $w_0$CDM models (for which $\wLJ$ is $w_0$), we find that $G$ and $dD/da$ agree closely with their exact solutions of \EQEQ{e:G_w0cdm}{e:dDda_w0cdm} for all $a\leq1$ and $w_0\leq -1/3$.

\subsection{Linear perturbation theory for neutrinos}
\label{sub:bkg:nupt}

We are only concerned here with linear perturbation theory in the sub-horizon case, so that questions of gauge choice may be set aside.  The standard Eulerian perturbative treatment of neutrinos, studied in Ref.~\cite{Ma:1995ey}, is discussed first, followed by the alternative Lagrangian-momentum approach of Refs.~\cite{Dupuy:2013jaa,Dupuy:2014vea,Chen:2020bdf}.  Each approach tracks a single neutrino four-momentum or four-velocity at once, with the total neutrino density contrast obtained by integrating over them.  Afterwards, we discuss the clustering and free-streaming limits which characterize neutrinos in both approaches, as well as a set of approximations interpolating between these limits.  All of these discussions easily generalize from neutrinos to arbitrary HDM, with appropriate substitutions for the particle mass and distribution function.


First, consider Eulerian perturbation theory, which tracks the motion of neutrinos through fixed coordinates in their six-dimensional phase space.  Our starting point is the Boltzmann equation in the sub-horizon, non-relativistic limit:
\begin{equation}
  \frac{df}{ds}
  = \frac{\partial f}{\partial s}
  + \frac{\vec p}{m_\nu} \cdot\vec\nabla_{\vec x} f
  - a^2 m_\nu \vec\nabla_{\vec x} \Phi \cdot \vec\nabla_{\vec p} f
  = 0.
  \label{e:eom_eul}
\end{equation}
At the time of neutrino decoupling, the distribution function $f$ is well-approximated by a spatially-uniform function $\bar f(p)$ of the magnitude of the momentum alone.  At later times, it develops perturbations, and we may write $f(a,\vec x, \vec p) = {\bar f}(p) + \delta\!f(a,\vec x, \vec p)$.

Since $\Phi$ is a first-order quantity in perturbation theory, we keep only the homogeneous part $\bar f$ of $f$ in the $\Phi \cdot \vec\nabla_{\vec p} f$ term.  Furthermore, since $\bar f$ is independent of $s$ and $\vec x$, the $\partial f /\partial s$ and $\vec\nabla_{\vec x} f$ terms both pick out the perturbation $\delta\!f$, resulting in the linear equation
\begin{equation}
  \frac{\partial \delta\!f}{\partial s}
  + \frac{\vec p}{m_\nu} \cdot\vec\nabla_{\vec x} \delta\!f
  = a^2 m_\nu \vec\nabla_{\vec x} \Phi \cdot \vec\nabla_{\vec p} {\bar f}.
  \label{e:eom_eul_lin}
\end{equation}
If the initial superconformal time $\sinit$ is chosen to be early enough that $\delta\!f$ may be neglected there, then \EQ{e:eom_eul_lin} has the formal solution
\begin{equation}
  \delta\!f(s,\vec k, \vec p)
  =
  \left( i m_\nu \vec k \cdot \vec \nabla_{\vec p} {\bar f} \right)
  \int_{\sinit}^s ds' a(s')^2 \Phi(s',\vec k)
  \exp\left( -\frac{i \vec k \cdot \vec p}{m_\nu} (s-s') \right).
  \label{e:deltaf_eul_lin_formal}
\end{equation}

Finally, note that \EQ{e:deltaf_eul_lin_formal} depends upon $\vec p$ only through its magnitude $p$ and its angle with $\vec k$, whose cosine is $\mu = \hat k \cdot \hat p$.  Averaging over $\mu$, we obtain the monopole perturbation $\delta\!f_0(s,\vec k,p)$.  As we are only interested in the monopole, henceforth we  use $\dfmu$ to represent $\delta\!f_0$:
\begin{eqnarray}
  \dfmu\!(s, \vec k, p)
  &:=&
  \delta\!f_0(s, \vec k, p)
  =
  \frac{k m_\nu}{p} \dfdlnp
  \int_{\sinit}^s ds' a(s')^2 \Phi(s',\vec k)
  W\!\left[\frac{k p (s-s')}{m_\nu}\right]
  \label{e:deltaf_eul_lin_ell0}
  \\
  W(X) &:=& \frac{\sin(X)}{X^2} - \frac{\cos(X)}{X}
  = -\frac{d}{dX} \left(\frac{\sin(X)}{X}\right).
\end{eqnarray}
This discussion generalizes to an arbitrary HDM species, provided that it decouples early enough that its perturbations may be neglected at that time.


Next, we consider an alternative approach to neutrino perturbation theory by Refs.~\cite{Dupuy:2013jaa,Dupuy:2014vea}, which treats neutrinos as Lagrangian in four-momentum-space.  That is, the clustering of neutrinos is studied as a function of their position and {\em initial} four-momentum, which is purely thermal, as neutrinos' initial velocity inhomogeneities may be neglected.  Since neutrinos with lower initial thermal momenta are more easily captured by collapsed structure, this Lagrangian-momentum approach provides useful insights into neutrino clustering.

Once again working under the assumption of neutrinos that are spatially homogeneous at the time of decoupling, we may decompose the four-momentum $P_\mu = m_\nu U_\mu$ into the sum of a spatially-homogeneous part $m_\nu {\bar U}_\mu$ and an inhomogeneous part $m_\nu \delta U_\mu$ which is negligible at decoupling.  Conveniently, the homogeneous lower-index four-velocity ${\bar U}_\mu$ has time-independent spatial components $\vec u := ({\bar U}_1, {\bar U}_2, {\bar U}_3)^T$, and its time component ${\bar U}_0 = -\sqrt{u^2 + a^2}$ is determined by the magnitude of $\vec u$.

A central insight of Refs.~\cite{Dupuy:2013jaa,Dupuy:2014vea} is that the neutrino distribution function $\bar f(m_\nu \bar U_\mu)$ at decoupling may be approximated by discretizing it into a collection of $\vec u_\alpha$.  Once a subset of the total neutrino population has been assigned to a $\vec u_\alpha$, the homogeneous component of its four-velocity, which is its initial four-velocity, is completely determined; these neutrinos cannot subsequently switch to a different $\vec u$.  Thus each such subset behaves as a separate fluid, obeying its own set of fluid equations, and interacting with other fluids only through the gravitational potentials.

References~\cite{Chen:2020bdf,Chen:2022dsv} derived the sub-horizon, non-relativistic case of the fluid equations obeyed by each sub-population as well as a formal integral solution:
\begin{eqnarray}
  \frac{\partial}{\partial s} \delta(a,k,u,\mu)
  &=&
  -iku\mu \delta(a,k,u,\mu) + a\Hc \theta(a,k,u,\mu)
  \label{e:eom_delta_a_k_u_mu}
  \\
  \frac{\partial}{\partial s} \theta(a,k,u,\mu)
  &=&
  -\left(1+\frac{d\log \Hc}{d\log a}\right)a\Hc \theta(a,k,u,\mu)
  - i k u \mu \theta(a,k,u,\mu) - \frac{ak^2}{\Hc}\Phi
  \label{e:eom_theta_a_k_u_mu}
  \\
  \delta_{\ell}(a,\vec k,u)
  &=&
  -(2\ell+1)k^2 \int_{\sinit}^{s(a)} ds'
  a(s')^2 (s-s') \Phi(a(s'), \vec k)
  j_\ell\!\left(uk(s-s')\right).
  \label{e:delta_ell_formal}
\end{eqnarray}
Here, $\mu := \hat u \cdot \hat k$ is the cosine of the angle between $\vec u$ and the Fourier wave number $\vec k$; $\Phi(a,k)$ is the gravitational potential; $\delta(a,k,u,\mu)$ is the density contrast of the sub-population characterized by $\vec u$, which we expand in multipoles as $\delta(a,k,u,\mu) = \sum_\ell (-i)^\ell \PLeg_\ell(\mu) \delta_\ell(a,k,u)$ with $\PLeg_\ell$ the Legendre polynomial; and $\theta(a,k,u,\mu) = -\nabla^i \delta U_i / (a\Hc)$ is the dimensionless velocity divergence.  Note that the angle of $\vec u$ affects clustering only through $\mu$.  For fixed $k$ and $\mu$, all $\vec u$ with the same magnitude $u$ obey the same equations of motion, so we refer to all such sub-populations collectively as a single ``flow'' specified by the flow speed $u$.

Of particular interest is the monopole solution $\ell=0$, which simplifies since $X j_0(X) = \sin(X)$.  Since we will not use the higher multipoles henceforth, we define 
\begin{equation}
  \delta(a,k,u) := \delta_0(a,k,u)
  = -\frac{k}{u} \int_{\sinit}^{s(a)} ds'
  a(s')^2  \Phi(a(s'),\vec k) \sin(uk(s-s')).
  \label{e:delta_lag_lin_ell0}
\end{equation}
Numerical solutions of this equation were considered further in Ref.~\cite{Upadhye:2025glk}.  Qualitatively, a higher $u$ at fixed $k$ makes the integrand more oscillatory, hence $\delta(a,k,u)$ less sensitive to the local gravitational potential, as expected of a freely-streaming particle.  Reference~\cite{Upadhye:2024ypg} showed that this Lagrangian description readily generalizes to arbitrary HDM, and even an effective HDM representing multiple species at once.


Note that the integrands in \EQEQ{e:deltaf_eul_lin_ell0}{e:delta_lag_lin_ell0} look similar, with the $W(X)$ function in \EQ{e:deltaf_eul_lin_ell0} replaced by $\sin(X)$ in \EQ{e:delta_lag_lin_ell0}.  In the clustering regime, corresponding to small $k$ and large length scales, the arguments of $W(X)$ and $\sin(X)$ become small, $W(X) \approx X/3$ and $\sin(X) \approx X$.  Thus both integrands are proportional to $a(s')^2 \Phi(a(s'),\vec k) (s-s')$.  Looking first at \EQ{e:delta_lag_lin_ell0}, we see that $\delta \rightarrow -k^2 \int ds' a(s')^2 \Phi(s',\vec k) (s-s')$.  Differentiating twice with respect to $s$ leads to $\partial^2\delta / \partial s^2 = -a^2 k^2 \Phi$, precisely the equation of motion obeyed by the linear CDM density contrast, hence the total matter contrast.  Thus, long after neutrinos have decoupled and become non-relativistic, we expect that $\dnu \rightarrow \dm$ as $k \rightarrow 0^+$.  Keeping track of constant factors, we also see that $\dfmu \rightarrow - \tfrac{1}{3} \dfdlnp \dm$ in the clustering regime.

In the free-streaming regime, at large $k$, neutrinos stream past  potential wells, clustering considerably more weakly than CDM.  Integrating \EQ{e:delta_lag_lin_ell0} by parts shows that
\begin{equation}
  \delta(a,k,u)
  =
  -\left[a(s')^2 \Phi(s',k) \tfrac{\cos(uk(s-s'))}{u^2}\right]_{\sinit}^s
  + \int_{\sinit}^{s(a)} ds' \tfrac{\cos(uk(s-s'))}{u^2}
  \tfrac{\partial}{\partial s'}(a(s')^2 \Phi(s',k)).
  \label{e:delta_a_k_u_int_by_parts}
\end{equation}
The integral is dominated by late times, $s-s' \lesssim 1/(uk)$, while $\partial (a^2\Phi) / \partial s' \sim \Hc a^2 \Phi$.  Thus when $\Hc / (uk) \ll 1$, that is, when $k \gg \Hc / u$, the integral will be small compared with the first term on the right hand side.  The first term, in turn, reduces to $-a(s)^2 \Phi(s,k) / u^2$ if we choose $\sinit$ sufficiently early that $a(\sinit) \ll a(s)$, which we are free to do.  Using Poisson's equation, $-k^2 \Phi = \frac{3}{2}\Om \Hc^2 \dm$, we may rewrite the free-streaming-regime density contrast as
\begin{equation}
  \frac{\delta(a,k,u)}{\dm(a,k)} \rightarrow \frac{\kfs(a,u)^2}{k^2}
  \quad\mathrm{where}\quad
  \kfs(a,u)^2 := \frac{3\Om \Hc^2 a^2}{2u^2} = \frac{3\Omo \Hco^2 a}{2u^2}.
  \label{e:def_kfs_lag}
\end{equation}
Here, $\kfs$ is the free-streaming wave number, which separates the clustering regime, $k \ll \kfs$, from the free-streaming regime, $k \gg \kfs$.  Deep inside the free-streaming regime, neutrinos cluster far more weakly than the total matter.

Here, and henceforth, let $x := k / \kfs$.  We have seen that $\delta(a,k,u)/\dm(a,k)$ approaches $1$ at low $x$ and $x^{-2}$ at high $x$.  The ``Generalized Super-Easy'' (GSE) fit of Refs.~\cite{Ringwald:2004np,Wong:2008ws,Chen:2020kxi,Pierobon:2024kpw} uses a simple rational function to interpolate between these:
\begin{equation}
  \delta(a,k,u) / \dm(a,k) \approx 1 / (1 + x + x^2)
  \quad\mathrm{where}\quad
  x(a,u) = k / \kfs(a,u).
  \label{e:gse_lag}
\end{equation}
Those references showed it to be accurate to about $5\%-10\%$ for standard $\Lambda$CDM models at late times, with the greatest errors in the interpolation region $x \sim 1$.

The free-streaming limit in the Eulerian case proceeds similarly, with the window function $W(X) = -d (\sin(X)/X) / dX$ replacing $\sin(X) = -d \cos(X)/X$:
\begin{eqnarray}
  \dfmu
  &=&
  \frac{m_\nu^2}{p^2} \dfdlnp
  \left[
    \left. a(s')^2 \Phi(s',k)
    \frac{\sin\!\!\left(\!\frac{kp(s-s')}{m_\nu}\!\right)
    }{\frac{kp(s-s')}{m_\nu}}\right|_{\sinit}^s
    - \int_{\sinit}^2 ds' \frac{d(a^2\Phi)}{ds'}
    \frac{\sin\!\!\left(\!\frac{kp(s-s')}{m_\nu}\!\right)}{\frac{kp(s-s')}{m_\nu}}
    \right]\qquad
  \label{e:eul_int_by_parts}
  \\
  \frac{\dfmu}{\dm}
  &\rightarrow& -\dfdlnp \frac{\kfs(a,p)^2}{k^2}
  \quad\mathrm{where}\quad
  \kfs(a,p)^2 := \frac{3\Om \Hc^2 a^2 m_\nu^2}{2p^2}
  = \frac{3\Omo\Hco^2m_\nu^2 a}{2p^2}
  \label{e:def_kfs_eul}
\end{eqnarray}
where the second line was obtained by discarding the integral in the first and by choosing $\sinit$ early enough that $a(\sinit) \ll a(s)$, as above.  References~\cite{Chen:2020kxi,Pierobon:2024kpw} find the interpolating function
\begin{equation}
  \frac{\dfmu}{\dm} \approx -\dfdlnp \,\frac{1}{3 + 2\sqrt{3}x + x^2}
  \quad\mathrm{where}\quad
  x(a,p) = k / \kfs(a,p)
  \label{e:gse_eul}
\end{equation}
to be accurate to $\lesssim 10\%$ for standard $\Lambda$CDM models at late times.  Our results in both the Eulerian and Lagrangian-momentum cases may easily be generalized to an arbitrary HDM species by replacing $m_\nu$ and $\bar f$  by the HDM mass and  distribution function, respectively.

The original Super-Easy fit of Ref.~\cite{Ringwald:2004np} fit the ratio of the integrated total neutrino density contrast $\dnu(a,k)$ to that of the matter:
\begin{equation}
  \frac{\dnu(a,k)}{\dm(a,k)} \approx \frac{1}{(1+k/\kfs(a))^2}
  \quad\mathrm{where}\quad
  \kfs(a)^2 := \frac{3\Omo \Hco^2 a}{2c_{\nu,0}^2},
  \quad
  c_{\nu,0}^2 := \frac{3 \zeta(3) \Tnuo^2}{2 \log(2) m_\nu^2}.
  \label{e:fit_se}
\end{equation}
Although this does not arise automatically from an integration of either of \EQEQ{e:gse_lag}{e:gse_eul}, it is an interpolation between the same clustering and free-streaming limits for the total neutrino fluid.  Here, $c_{\nu,0}$ is a neutrino sound speed, and the resulting free-streaming scale $\kfs(a)$ is an average free-streaming scale describing the entire neutrino population.

\subsection{Effective HDM and neutrino hierarchies}
\label{sub:bkg:ehdm}

References~\cite{Bayer:2020tko,Upadhye:2024ypg} demonstrated that an arbitrary collection of HDM species, even at times when one or more are relativistic, may be represented using a single effective hot dark matter (EHDM) species with an appropriately-defined distribution function.  With $m_\hdm^{(s)}$, $T_{\hdm,0}^{(s)}$,  $g_\hdm^{(s)}$, and ${\bar f}_\hdm^{(s)}(q)$ the mass, temperature, multiplicity, and distribution of HDM species $s$, respectively, and $q$ the momentum-to-temperature ratio, the effective distribution function is
\begin{equation}
  {\bar f}_\ehdm(q)
  =
  \sum_{{\rm species}~s}
  g_\hdm^{(s)} \left(\frac{m_\hdm^{(s)}}{m_\ehdm}\right)^4
  {\bar f}_\hdm^{(s)}\!\!
  \left(q \frac{T_{\ehdm,0}}{T_{\hdm,0}^{(s)}} \frac{m_\hdm^{(s)}}{m_\ehdm}\right).
  \label{e:f_edhm}
\end{equation}
Here, $m_\ehdm$ and $T_{\ehdm,0}$ are the mass and redshift-zero temperature of the EHDM.  Reference~\cite{Upadhye:2024ypg} chooses $m_\ehdm$ to be the density-weighted average of the HDM masses and $T_{\ehdm,0}$ the mass-weighted average of the HDM temperatures.

This EHDM description is a consequence of the Equivalence Principle, by which two decoupled HDM particles beginning in the same position $x^\mu$ and velocity $U_\mu$ will evolve the same way, regardless of mass.  As a result, different combinations of the same set of Lagrangian $u$ can be used to represent different HDM species, or all HDM at once.  Reference~\cite{Upadhye:2024ypg} demonstrates that a set of $10-20$ appropriately-chosen $u$ is sufficient to represent two neutrino species of different masses and an axion, whose masses span a factor of $\approx 25$.

Thus we may represent three neutrino species with a normal ordering (NO) or inverted ordering (IO) of masses using a single effective neutrino species.  In this work, we are concerned with the clustering of individual HDM $u$ or $p$.  In our example neutrino models, we will assume a degenerate mass ordering (DO) until Sec.~\ref{sec:gen}.  Using a modified distribution function, our results can easily be generalized to other mass orderings or HDM models.

\section{Exact Einstein-de Sitter solutions}
\label{sec:eds}

The GSE approximation of Sec.~\ref{sub:bkg:nupt} fit the ratio of the neutrino density contrast, for a single momentum, to that of the total matter.  There are also exact solutions for the neutrino-to-matter ratio, in a simplified approximation to the EdS case which we define in Sec.~\ref{sub:eds:test-hdm}.  Starting from the formal solutions of Sec.~\ref{sub:bkg:nupt}, we derive these exact EdS solutions in Secs.~\ref{sub:eds:eul},~\ref{sub:eds:lag}.  We then proceed to study their behavior in the clustering and free-streaming limits.  Although the EdS solutions, used as approximations to the $\Lambda$CDM neutrino-to-matter ratios, have larger maximum errors than the GSE fits, they are more accurate in each of these limits.  The subsequent Section will use these exact solutions to develop an even more accurate fit.

\subsection{EdS-test-HDM case}
\label{sub:eds:test-hdm}

Exact solutions for the Eulerian distribution function perturbation $\dfmu(a,k,p)$ and the Lagrangian-momentum density contrast $\delta(a,k,u)$ can be found in a particular case, which we call the EdS-test-HDM case, that is defined as follows.
\begin{enumerate}
\item The HDM is non-relativistic at all times considered.
\item HDM behaves as a test fluid, that is, it does not enter into the
  gravitational potential determining the growth of the total matter
  density contrast $\dm(a,k)$.
\item The universe is described by a linear Einstein-de Sitter cosmology, that
  is, $\Om(a) = 1$, with no curvature, dark energy, radiation, or other species,
  and non-linear corrections to the growth are negligible, so that
  $\dm(a,k) \propto a$.
\end{enumerate}

Section~\ref{sub:bkg:nupt} provided formal solutions for $\dfmu(a,k,p)$ and $\delta(a,k,u)$ given a gravitational potential $\Phi(a,k) \propto -\Om \Hc^2 \dm / k^2$. In the EdS-test-HDM case, these are no longer merely formal solutions.  Since $\dm(a,k) \propto D(a) = a$, given any $a$ we have $\dm(a',k) = \dm(a,k) a' / a$, of which $\dm(a,k)/a$ may be pulled outside the integrals.  Then, if the remaining integrals may be carried out exactly, the results will be closed expressions for $\dfmu(a,k,p)/\dm(a,k)$ and $\delta(a,k,u)/\dm(a,k)$ which are exactly applicable in the EdS-test-HDM case.

\subsection{Eulerian perturbation theory}
\label{sub:eds:eul}

In this section, we switch from superconformal time to a related variable, $y = - p k s / m_\nu$ in the Eulerian case and $y = - u k s$ in the Lagrangian-momentum case.  This has the advantages of being positive and dimensionless.  The superconformal time $s$ is  useful for studying particles with redshifting non-relativistic velocities $\vec v = \vec u / a$, because the distance travelled simplifies: $\int d\tconf \, \vec v(\tconf) = \int d\tconf  \, \vec u / a = \vec u \int ds = \vec u \Delta s$.  Since $k$ is inversely proportional to the wavelength, $y$ is thus a dimensionless path length measured in units of wavelength.

Rewriting \EQ{e:deltaf_eul_lin_ell0} as an integral over $y$, we have
\begin{eqnarray}
  \dfmu(a,k,p)
  &=&
  \frac{m_\nu^2}{k^2 p^2} \dfdlnp
  \int_{y(a)}^{\yin} dy' a(y')^2 k^2 \Phi(y',k) W(y'-y)
  \nonumber\\
  &=&
  -\frac{1}{a x^2} \dfdlnp
  \int_{y(a)}^{\yin} dy' a(y') \dm(y',k) W(y'-y)
\end{eqnarray}
where the second equality uses Poisson's equation as well as \EQ{e:def_kfs_eul} for $\kfs$.  In the general linear case, we may write $\dm(a',k) = \dm(a,k) D(a',k) / D(a,k)$, allowing us to pull $\dm(a,k)$ outside the integral.  We are therefore left with the ratio
\begin{equation}
  \frac{\dfmu(a,k,p)}{\dm(a,k)}
  =
  -\dfdlnp \frac{\psi(a,k,y)}{x^2},
  \quad
  \psi(a,k,y)
  :=
  \int_y^{\yin} dy' \frac{a(y') D(a(y'),k)}{a(y) D(a(y),k)} W(y'-y).
  \label{e:psi}
\end{equation}
Recall that $W(y) = \sin(y)/y^2 - \cos(y)/y$, while $\sin(y'-y)/(y')^n$ and  $\cos(y'-y)/(y')^n$ may be integrated exactly over $y'$ for any integer $n$ through the repeated use of integration by parts.

In the EdS-test-HDM case, $D(a') = a'$, reducing the integrand of $\psi$ to $(y/y')^4 W(y'-y)$.  Further, since $y \propto s \propto a^{-1/2}$, the $a\rightarrow 0^+$ limit corresponds to $\yin \rightarrow \infty$.   We take this limit here.  Then the exact solution, in terms of the sine integral $\Si(y)$ and cosine integral $\Ci(y)$, is
\begin{eqnarray}
  \psieds(x)
  &=&
  -x^2 + 4 - \sqrt{\frac{2}{3}}\frac{\pi}{x}
  + 4(x^2 - 1)
  \left[\sin(x\sqrt{6})\left(\Si(x\sqrt{6})
    -\frac{\pi}{2}\right) + \cos(x\sqrt{6})\Ci(x\sqrt{6})\right]
  \ldots\nonumber\\
  &~&
  +  \left(\!-x^3\sqrt{6} + 2\sqrt{6} x - \!\frac{2\sqrt{2}}{x\sqrt{3}} \right)
  \left[\cos(x\sqrt{6})\left(\Si(x\sqrt{6})-\frac{\pi}{2}\right)
    - \sin(x\sqrt{6})\Ci(x\sqrt{6})\right]
  \label{e:psieds}
\end{eqnarray}
where we have used the fact that, in EdS, the simple form of the superconformal time implies that $y = x \sqrt{6}$.
Thus $\dfmu/\dm$ reduces to a function of $x$ alone.

In the clustering limit, where HDM clusters like CDM, the  $\dfmu/\dm$ ratio approaches a constant.  We find from \EQ{e:psieds} that
\begin{equation}
  \frac{\dfmu(a,k,p)}{\dm(a,k)}
  =
  -\frac{1}{3} \dfdlnp
  - \frac{6}{5} \dfdlnp x^2 \Ci(x\sqrt{6})
  + \frac{1}{5} \dfdlnp x^2 + {\mathcal O}(x^3).
  \label{e:df_dm_eul_clus}
\end{equation}
Consider the case of a single neutrino species, for which $\bar f$ is the Fermi-Dirac distribution.  Integrating the first term of \EQ{e:df_dm_eul_clus} over momenta and dividing by the neutrino number density, we find $\dnu/\dm \rightarrow 1$, precisely as expected in the clustering limit.

Meanwhile, in the free-streaming limit $x \rightarrow \infty$, asymptotic expansions of the sine and cosine integrals may be used to demonstrate that
\begin{equation}
  \psieds
  \rightarrow
  1 - \frac{2\pi}{y} + \frac{20}{y^2} - \frac{280}{y^4}  + {\mathcal O}(y^{-6})
  =
  1 - \sqrt{\frac{2}{3}}\frac{\pi}{x} + \frac{10}{3x^2} - \frac{70}{9x^4}
  + {\mathcal O}(x^{-6}).
\end{equation}
Substituting this into the expression for $\dfmu/\dm$ in \EQ{e:psi}, we see to leading order that it approaches $-x^{-2} d{\bar f}/d\log p$, consistent with \EQ{e:def_kfs_eul}.

\subsection{Lagrangian momenta}
\label{sub:eds:lag}

Beginning with \EQ{e:delta_lag_lin_ell0} and changing from $s$ to $y = -uks$, we have
\begin{eqnarray}
  \delta(a,k,u)
  &=&
  -\frac{1}{k^2 u^2} \int_{y(a)}^{\yin} a(y')^2 k^2 \Phi(y',k) \sin(y'-y)
  \nonumber\\
  &=&
  \frac{1}{a x^2} \int_{y(a)}^{\yin} a(y') \dm(a(y'),k) \sin(y'-y).
  \label{e:delta_a_k_u_int_y}
\end{eqnarray}
Once again writing $\dm(a',k) = \dm(a,k) D(a',k) / D(a,k)$ in order to pull $\dm(a,k)$ outside the integral, we find that
\begin{equation}
  \frac{\delta(a,k,u)}{\dm(a,k)}
  =
  \frac{1}{x^2}
  \int_{y(a)}^{\yin} dy'  \frac{a(y') D(a(y'),k)}{a D(a,k)} \sin(y'-y).
\end{equation}
Note the similarity of the integrand to that of $\psi(y)$ in \EQ{e:psi}; the only difference here is a simpler window function, $\sin(y'-y)$.

Once again, in the EdS case, we take the $\yin \rightarrow \infty$ limit setting the initial scale factor to zero.  The exact solution for $D(a,k) = a$ is
\begin{equation}
  \left[ \frac{\delta(a,k,u)}{\dm(a,k)}\right]_{\rm EdS}
  =
  1 + 6x^2 \cos(x\sqrt{6}) \Ci(x\sqrt{6})
  + 6x^2 \sin(x\sqrt{6}) \left[\Si(x\sqrt{6}) -\frac{\pi}{2}\right].
  \label{e:dhdm_dm_eds_lag}
\end{equation}
As above, $y = x \sqrt{6}$ in the EdS case.  The leading-order clustering-limit solution $\delta/\dm \rightarrow 1$ is evident, while the leading-order free-streaming limit $\delta/\dm \rightarrow 1/x^2$ follows from the asymptotic expansions of $\Si(y)$ and $\Ci(y)$.

Finally, consider the next-to-leading-order (NLO) term in the clustering limit: $6 x^2 \Ci(x\sqrt{6})$. At low $x$, $\Ci(y) \approx \log(y) + \gammaEM$, with $\gammaEM \approx 0.577$ the Euler-Mascheroni constant.  Thus the NLO term splits into one term $\propto x^2 \log(x)$ and another $\propto x^2$.  The same is true of the NLO terms of $\dfmu / \dm / \dfdlnp$ in \EQ{e:df_dm_eul_clus}.  The presence of this logarithmic term significantly complicates attempts at higher-order corrections in the clustering regime to the Super-Easy fits of Sec.~\ref{sub:bkg:nupt}.  While there is no Taylor series representation of $x^2 \log(x)$ at $x=0$, this term is comparable in magnitude to other ${\mathcal O}(x^2)$ terms for $x \sim 0.1$.

\begin{figure}[tb]
  \includegraphics[width=75mm]{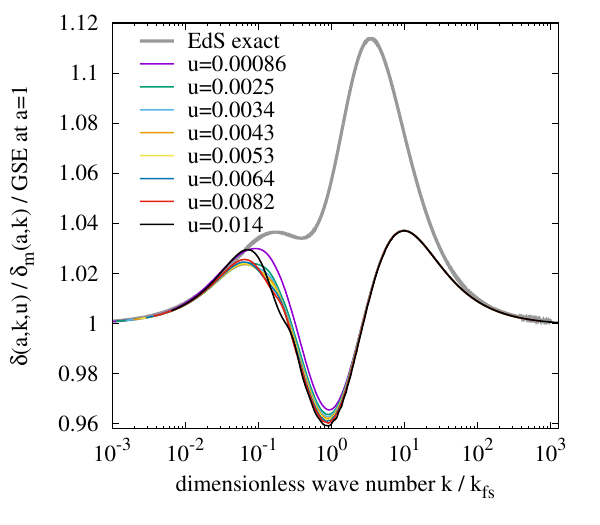}
  ~\includegraphics[width=75mm]{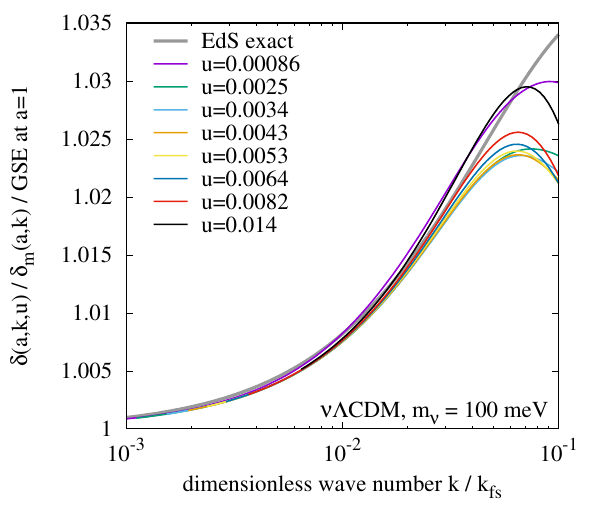}%
  \caption{
    Exact EdS solution of \EQ{e:dhdm_dm_eds_lag} for $\delta(a,k,u)/\dm(a,k)$
    (thick gray line) vs.\,\,several Lagrangian flows, for a $\nu\Lambda$CDM
    model with $\Omo=0.3$ and $\Mnu=300$~meV ($m_\nu=100$~meV) at $a=1$ (thin
    colored lines).  Each $\delta/\dm$ is shown relative to the GSE fit of
    \EQ{e:gse_lag}.
    (Left)~The different $\nu\Lambda$CDM $\delta / \dm$ ratios agree with one
    another at the $\approx 1\%$ level, while the EdS solution substantially
    overestimates them.
    (Right)~At $x \lesssim 0.1$, the EdS solution nevertheless predicts the
    $\approx 3\%$ rise in $\delta / \dm$ above the GSE fit,
    showing that higher-order EdS corrections are accurate there.
    \label{f:lag_eds_vs_nulcdm}
  }
\end{figure}

The inclusion of this non-perturbative $x^2 \log(x)$ term in both \EQEQ{e:psieds}{e:dhdm_dm_eds_lag} is one of our main motivations for seeking a fit relative to these EdS solutions.  The reader may worry that these EdS NLO terms would be inapplicable to very different cosmological models.  However, Figure~\ref{f:lag_eds_vs_nulcdm} demonstrates that the higher-order terms of \EQ{e:dhdm_dm_eds_lag} do accurately predict deviations from the GSE fit for $x$ up to $0.1$ for a $\nu\Lambda$CDM model very different from EdS.

\section{SE-EdS fitting functions}
\label{sec:fit}

Having derived the exact EdS-test-HDM solutions in the previous section, we proceed to use them to improve upon the accuracy of the GSE fit.  Since the non-perturbative clustering-limit terms discussed above complicate rigorous expansions, we instead take a phenomenological approach here.  In Sec.~\ref{sub:fit:3limits}, we consider three limits in which the HDM-to-matter ratio for general cosmologies ought to approach the EdS case. Sections~\ref{sub:fit:eul},~\ref{sub:fit:lag} use these limits to parameterize the differences in HDM-to-matter ratios between EdS and more general $w_0$CDM cosmologies, and, hence, fitting functions for $w_0$CDM models.  These fits are shown to be accurate to $<3\%$ for a broad range of cosmological parameters. 

Readers wishing only to paint neutrinos into N-body simulations, with a minimum of technical details, may skip directly to Sec.~\ref{sub:fit:paint}.  There, we show how to obtain the neutrino density contrast and the corrected gravitational potential, starting from the density contrast of the simulated particles, which represent only the CDM and baryons.  We have kept the discussion self-contained for ease of use.

\subsection{Three EdS limits}
\label{sub:fit:3limits}

Rather than seeking fits to the HDM-to-matter ratios, we fit the ratios of these ratios to their exact EdS solutions, \EQEQ{e:psieds}{e:dhdm_dm_eds_lag}, from Sec.~\ref{sec:eds}:
\begin{equation}
  \RE(a,k,p):=
  \frac{\dfmu(a,k,p) / \dm(a,k)
  }{\left[\dfmu(a,k,p) / \dm(a,k)\right]_{\rm EdS}},
  \qquad
  \RL(a,k,u) :=
  \frac{\delta(a,k,u)/\dm(a,k)}{\left[\delta(a,k,u)/\dm(a,k)\right]_{\rm EdS}}.
  \label{e:RE_RL}
\end{equation}
Figure~\ref{f:lag_eds_vs_nulcdm} compares $\delta(a,k,u)/\dm(a,k)$ for several Lagrangian flows to the exact EdS ratio of \EQ{e:dhdm_dm_eds_lag}.  Note several trends.  Firstly, when $\delta/\dm$ for the different flows are plotted as functions of $x$, they agree with one another at the percent level, suggesting a universal fitting function of $a$ and $x$ alone.  Separate dependences upon $k$ and $u$ can be approximated by a dependence upon the single variable $x$.  Secondly, the exact EdS $\delta/\dm$ overestimates each of these flows around $x \sim 1$, implying that a fit approximating $\RL$ there must be less than unity.  Since the flows agree with one another, the magnitude of this overestimate must be a function of the scale factor and cosmological parameters, but not $u$.  Thirdly, reiterating the observation made in Sec.~\ref{sub:eds:lag}, the NLO EdS terms in the $x \leq 0.1$ range are more accurate than the GSE fit of \EQ{e:gse_lag}, accurately predicting the $\approx 3\%$ rise above GSE.

Aside from $x\rightarrow 0^+$ and $x\rightarrow \infty$, we demand that our fitting functions, $\FE(a,x)$ for $\RE(a,k,p)$ and $\FL(a,x)$ for $\RL(a,k,u)$, approach unity in three specific limits:
\begin{enumerate}
\item $\Omo \rightarrow 1^-$, which,
  in our spatially-flat cases, implies that EdS is approached directly;
\item $a \rightarrow 0^+$, since our radiation-free approximation means that
  any model with $w$ bounded below zero approaches EdS at early times; and
\item $w_0 \rightarrow -\infty$ at fixed $\Omo$, which pulls the scale factor
  of matter-dark-energy equality arbitrarily close to unity, making the
  dark-energy-dominated era negligible.
\end{enumerate}

The third limit requires further discussion.  The scale factor of matter-dark-energy equality $\amd = r_0^{-1/(3w_0)}$, meaning that as $w_0\rightarrow -\infty$, $1-\amd \rightarrow \tfrac{1}{3w_0} \log(r_0) = |\log(r_0)| / |3w_0|$.  If $G \approx 1$ until $\amd$ and only falls after that, then we expect $1-G \sim \woai$
.  A more careful analysis applied to \EQ{e:G_w0cdm} reveals that, in the limit of small positive $(-w_0)^{-1}$, 
\begin{equation}
  1-G(a) \rightarrow \frac{2}{3} \woai
  \log\!\left(\frac{1+\Om(a)^{1/2}}{2\Om(a)^{1/2}}\right).
  \label{e:G_large_w0}
\end{equation}
Evidently, this is positive for all $0 < \Om(a) < 1$ and vanishes as $\woai \rightarrow 0^+$.  Due to the logarithmic term, it also vanishes in the limit $\Om(a) \rightarrow 1^-$, which is automatically approached as $a \rightarrow 0^+$.  Thus $1-G(a)$ vanishes in each of the three EdS limits identified above.  Our strategy for designing a fitting function, then, is to choose a rational function whose numerator and denominator become equal as $1-G(a)\rightarrow 0$.

\subsection{Eulerian \SEEdS{} fit}
\label{sub:fit:eul}

We find that $\RE(a,k,p)$ may be approximated using the remarkably simple fitting function

\noindent
\framebox[0.99\width][c]{
  \parbox[c]{153mm}{
    \begin{equation}
    \FE(a,x) = \frac{9+x^2}{9 + 3[1-G(a)]x + x^2} \approx \RE(a,k,p).
    \label{e:fit_eul}
    \end{equation}
  }
}

\noindent
Since this is a Super-Easy-like fit applied to the ratio $\RE$ to the EdS solution, we refer to this fit and its Lagrangian-momentum counterpart as the \SEEdS{} fits.  In order to test its accuracy, we begin with a fiducial $\nu\Lambda$CDM  model:
\begin{equation}
  \Omo=0.33,~
  \Obo=0.05,~
  \Ono=0.004,~
  h=0.66,~
  \sigma_8=0.83,~
  \ns=0.97,~
  w_0=-1.
  \label{e:fid_cosmo}
\end{equation}
We assume degenerate neutrino masses, implying three neutrino species with individual masses $m_\nu = 54.16$~meV.  Our reference computation, against which we compare \EQ{e:fit_eul}, is a direct numerical integration of $\psi$ in \EQ{e:psi} using the leading-order growth factor of Ref.~\cite{Upadhye:2025glk}, which applies a Super-Easy-like interpolation to  the $\approx 4 \fnu$ small-scale suppression of $\dm(a,k)$ by neutrinos.  We compare this numerically-integrated $\psi(a,k,x)$ to $\FE(a,x) \times \psieds(x)$.

\begin{figure}[tb]
  \hskip-2mm\includegraphics[width=79mm]{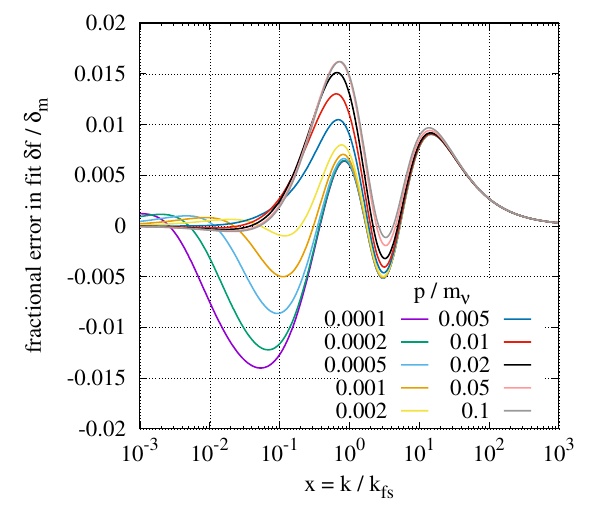}
  \includegraphics[width=79mm]{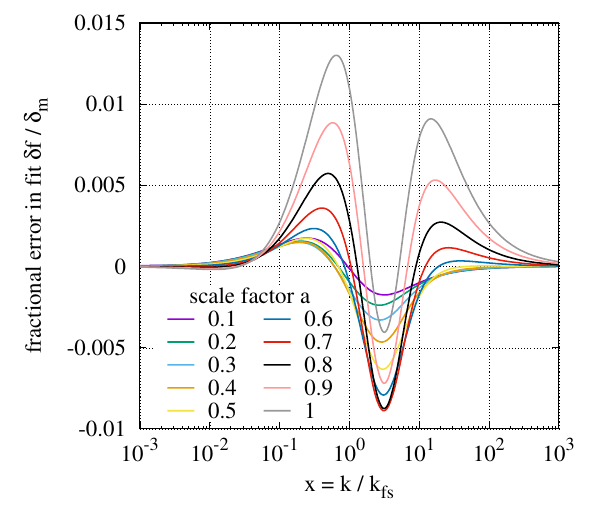}%
  \caption{
    Fractional error in the fit of \EQ{e:fit_eul} times the exact EdS solution
    of \EQ{e:psieds}, vs.\,\,direct numerical integration of $\psi(a,k,x)$
    in \EQ{e:psi}, for the fiducial model of \EQ{e:fid_cosmo}.
    (Left)~Variation of the momentum $p$ at $a=1$.
    (Right)~Variation of $a$ at $p/m_\nu = 0.01$.
    \label{f:eul_vary_p_a}
  }
\end{figure}

Figure~\ref{f:eul_vary_p_a} tests $\FE(a,x)\times \psieds(x)$ as an approximation to $\psi(a,k,x)$ for this fiducial model.  At all times, momenta, and wave numbers shown, errors are no more than $1.6\%$.  The plot on the left varies the neutrino momentum.  The $1\%-2\%$ scatter in the plot, consistent with Fig.~\ref{f:lag_eds_vs_nulcdm}, shows the limitation in approximating $\psi(a,k,x)$ for all $p$ using a single function of $a$ and $x$ alone.  The implication is that tuning the coefficients of \EQ{e:fit_eul} will not yield significant improvements.  Closer examination shows two regimes, a low-$p$ regime with $p/m_\nu \lesssim 0.001$ and maximum errors at $x \lesssim 0.1$, and a middle-to-high-$p$ regime with $p/m_\nu \gtrsim 0.001$ and maximum errors at $x \approx 1$.  Since maximum errors in both regimes are similar in magnitude, $\approx 1.5\%$, we focus on the case $p/m_\nu = 0.01$, approximately the mean momentum for our fiducial model, until we test the total $\dnu(a,k)$ at the end of Sec.~\ref{sub:fit:eul}.

Figure~\ref{f:eul_vary_p_a}~(Right) tests the fit of \EQ{e:fit_eul} with $p/m_\nu = 0.01$ for a range of scale factors.  Not surprisingly, errors fall at earlier times, as the cosmology approaches EdS.  We find errors $<0.5\%$ in magnitude everywhere for $a \leq 0.4$ and errors $<0.2\%$ for $a \leq 0.1$.  Since N-body simulations typically neglect the relativistic kinematics of neutrino particles, our fit will agree increasingly well with the simulated neutrino density contrast at earlier times.

\begin{figure}[tb]
  \hskip-2mm\includegraphics[width=79mm]{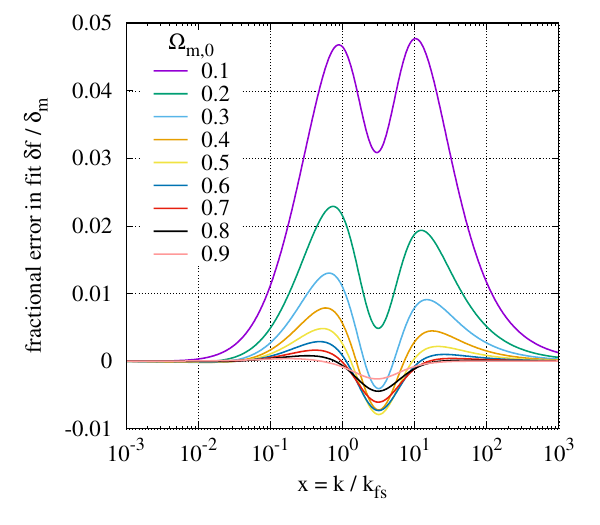}
  \includegraphics[width=79mm]{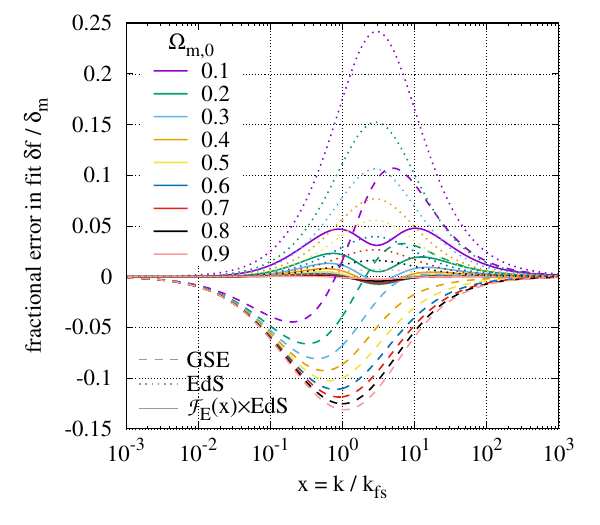}%
  \caption{
    Fractional error in the GSE, EdS, and \SEEdS{} $\FE(x)$ fits for 
    a range of $\Omo$, with all other parameters fixed to their 
    fiducial values of \EQ{e:fid_cosmo}, for $p/m_\nu=0.01$ and $a=1$.
    (Left)~$\FE(x)$ fit of \EQ{e:fit_eul}.
    (Right)~All three methods compared.
    \label{f:eul_vary_Om0}
  }
\end{figure}

Next, we deviate from the fiducial model.  Figure~\ref{f:eul_vary_Om0}~(Left) tests our $\FE(x)$ fit of \EQ{e:fit_eul} for a wide range of $\Omo$.  Surprisingly, the fit remains $\approx 5\%$ accurate even down to $\Omo=0.1$, and better than $3\%$ accurate for $\Omo \geq 0.2$.  Figure~\ref{f:eul_vary_Om0}~(Right) compares our fit $\psi \approx \FE \psieds$ to the GSE fit of Sec.~\ref{e:gse_eul} as well as the EdS approximation $\psi \approx \psieds$ of \EQ{e:psieds}.  The GSE fit is accurate at low matter densities, $\Omo \lesssim 0.3$, for which it was designed, but its errors exceed $10\%$ for $\Omo \geq 0.5$, implying early-time inaccuracy when it is used to approximate linear neutrinos in simulations.  The EdS approximation is accurate for $\Omo \gtrsim 0.4$, implying high accuracy at early times, but exceeds $10\%$ errors for values $\Omo \approx 0.3$ preferred by the data.  Our $\FE$ fit is more accurate than either GSE or EdS across the entire range of $\Omo$.

\begin{figure}[tb]
  \hskip-2mm\includegraphics[width=79mm]{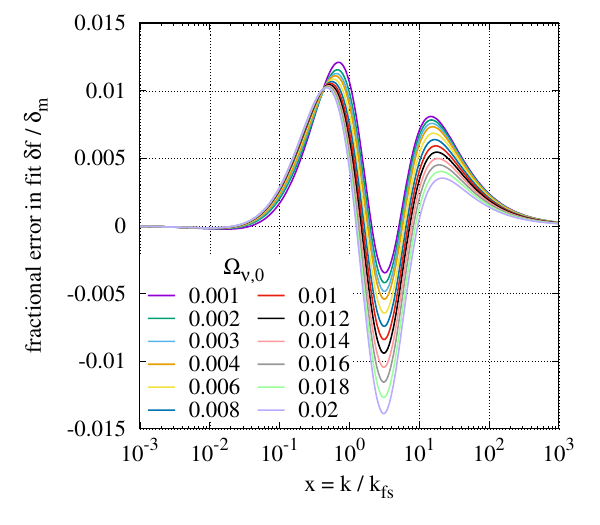}
  \includegraphics[width=79mm]{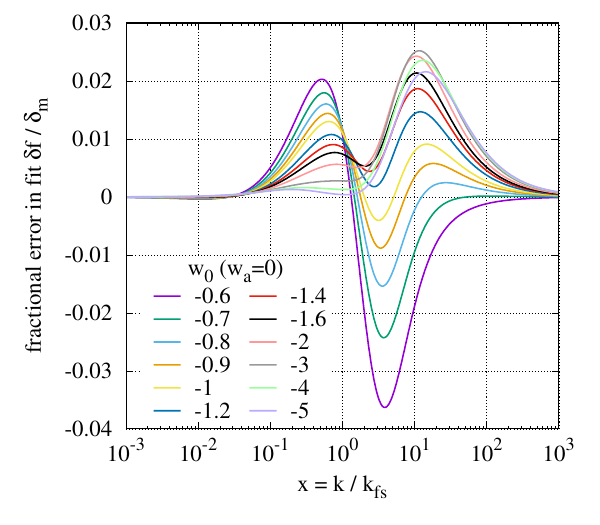}%
  \caption{
    Fractional error in the \SEEdS{} $\FE$ fit of \EQ{e:fit_eul} for 
    $p/m_\nu=0.01$ and $a=1$ as single parameters are varied about the 
    fiducial model of \EQ{e:fid_cosmo}.
    (Left)~Vary $\Ono$, continuing to assume three degenerate-mass species.
    (Right)~Vary the constant dark energy equation of state $w_0$.
    \label{f:eul_vary_Onu0_w0}
  }
\end{figure}

Figure~\ref{f:eul_vary_Onu0_w0} varies the neutrino fraction and the constant dark energy equation of state.  The neutrino fraction has only a small impact on the error in our method.  Meanwhile, Fig.~\ref{f:eul_vary_Onu0_w0}~(Right) shows two distinct behaviors.  As $w_0$ is increased above $-1$, the time period of dark energy domination increases.  For our fiducial $\Omo$, $\amd$ is $\approx 0.79$ for $w_0=-1$ and $\approx 0.62$ for $w_0=-0.5$.  Thus errors increase, exceeding $3\%$ between $w_0$ of $-0.7$ and $-0.6$.  In the other direction, errors increase to $\approx 2.5\%$ as $w_0$ is reduced to $-3$, but then decrease as it is further reduced.  This is due to the duration of dark energy domination decreasing, increasing the regime of validity of the EdS approximation.  Thus we find $<3\%$ accuracy for the range $w_0 \leq -0.7$, extending to arbitrarily negative $w_0$.

Thus far, we have compared our fit to the numerically-integrated $\psi(a,k,x)$, both within the context of our non-relativistic neutrino approximation.  We conclude by comparing the integrated total neutrino density contrast to the result of the \classcode{} code.  The relativistic-to-non-relativistic transition imprints a feature onto the neutrino density contrast, characterized by the transition scale factor $\anu$ and wave number $\knr$:
\begin{equation}
  \anu = \frac{7\pi^4 \Tnuo}{180 \zeta(3) m_\nu},
  \qquad
  \knr = \sqrt{\frac{7\pi^4 \log(2) \Omo \Hco^2 m_\nu}{180\zeta(3)^2 \Tnuo}},
\end{equation}
where $\zeta(z)$ is the Riemann zeta function; see Refs.~\cite{Bond:1980ha,Hu:1997vi}.  At a time $a \gg \anu$, the amplitude of this feature is $\sim \anu/a$.  We find that the feature peaks at a wavenumber $\sim \knr/3$, and that it can be fit by multiplying our non-relativistic-neutrino $\dnu$ by the function
\begin{equation}
  \Fnr(a,k) = 1 + \frac{(\anu/a) (3k/\knr)}{ 1 + (3k/\knr)^4}.
  \label{e:fit_r-nr}
\end{equation}
Integrating $\dfmu$ over momenta to obtain $\delta \rho_\nu = {\bar \rho}_\nu \dnu$, our expression for $\dnu/\dm$ is thus

\noindent
\framebox[0.99\width][c]{
  \parbox[c]{153mm}{
    \begin{eqnarray}
      \frac{\dnu(a,k)}{\dm(a,k)}
      &=&
      \frac{2}{3\zeta(3)} \Fnr(a,k)
      \int_0^\infty dq\,q^3 \frac{e^q}{(e^q+1)^2} {\mathcal G}_{\rm E}(a,C(a)q)
      \label{e:dnu_tot_eul}
      \\
      \mathrm{where}~
             {\mathcal G}_{\rm E}(a,x) &:=& \frac{1}{x^2} \FE(a,x) \psieds(x)
             \quad\mathrm{and}\quad
             C(a):=\sqrt{\frac{2}{3\Omo a}} \frac{k \Tnuo}{\Hco m_\nu}.
             \label{e:def_GE_Ca}
    \end{eqnarray}
  }
}

\noindent
Here, $q = p/T$, and for the Fermi-Dirac distribution, $\dfdlnp = q e^q (e^q+1)^{-2}$.  This integrated density ratio, and its Lagrangian counterpart to be derived in Sec.~\ref{sub:fit:lag}, are the main results of this Article.  The $x$-dependent function ${\mathcal G}_{\rm E}$ must be included in the integral because $x$ depends upon $p$, hence $q$, through the momentum-dependent free-streaming scale: $x = C(a) q$. Since $e^q (e^q+1)^{-2} \sim e^{-q}$ at large $q$, we integrate this numerically using Gauss-Laguerre quadrature.

\begin{figure}[tb]
  \begin{center}
    \includegraphics[width=140mm]{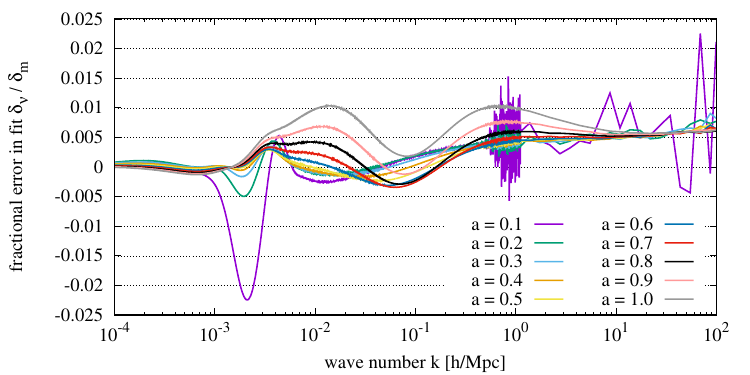}%
  \end{center}
  \caption{
    Comparison of the total neutrino density contrast of \EQ{e:dnu_tot_eul},
    divided by $\dm$, to that found using \classcode{} for the fiducial model
    of \EQ{e:fid_cosmo}.
    \label{f:eul_dnu_dm_vs_class}
  }
\end{figure}

Figure~\ref{f:eul_dnu_dm_vs_class} compares this $\dnu/\dm$ to that computed by the \classcode{} code.  Over most of the $a$ and $k$ range, errors are $\leq 1\%$, implying some cancellation of the $\approx 1.5\%$ errors of Fig.~\ref{f:eul_vary_p_a}~(Left).  Errors exceed $1\%$ in two separate regimes.  At $k \approx 0.00154~h$/Mpc$=\knr/3$, we see a residual error, associated with the relativistic-to-non-relativistic transition, which scales as $(\anu/a)^2$.  While this may be fit by adding higher-order terms in $\anu/a$ to $\Fnr$, we do not persue such a fit here.  The second regime with errors larger than $1\%$, $k \gtrsim 1~h$/Mpc, is due to a $1\%-2\%$ noise seen in the \classcode{} neutrino density contrast at $a=0.1$ which becomes worse at earlier times; our \EQ{e:dnu_tot_eul} is smooth in that region.  Setting aside this noise, we see agreement with \classcode{} to $\leq 1\%$ at all $k \geq \knr = 0.00463~h$/Mpc, and agreement to $< 2.5\%$ for all $a$ and $k$ shown.  We have thus confirmed the accuracy of the fit \EQ{e:fit_eul} relative to our own numerical calculations as well as the standard \classcode{} code.

\subsection{Lagrangian \SEEdS{} fit}
\label{sub:fit:lag}

Reference~\cite{Chen:2020kxi}, studying Eulerian and Lagrangian-momentum GSE approximations, observes:
\begin{equation}
  -\frac{3 \dfmu}{\dm \dfdlnp} \rightarrow
  \left\{
  \begin{array}{cl}
    1 & \mathrm{when}~x\rightarrow 0^+ \\
    3/x^2 & \mathrm{when}~x\rightarrow \infty \\
  \end{array}
  \right. ,
  \qquad
  \frac{\delta(a,k,u)}{\dm(a,k)} \rightarrow
  \left\{
  \begin{array}{cl}
    1 & \mathrm{when}~x\rightarrow 0^+ \\
    1/x^2 & \mathrm{when}~x\rightarrow \infty \\
  \end{array}
  \right. .
\end{equation}
That is, the Lagrangian-momentum free-streaming limit may be found from the Eulerian one through the substitution $x \rightarrow x\sqrt{3}$.  This suggests defining the Lagrangian-momentum \SEEdS{} fitting function as $\FL(a,x) = \FE(a,x\sqrt{3})$:

\noindent
\framebox[0.99\width][c]{
  \parbox[c]{153mm}{
  \begin{equation}
    \FL(a,x) = \frac{3+x^2}{3 + \sqrt{3}[1-G(a)]x + x^2} \approx \RL(a,k,u).
    \label{e:fit_lag}
  \end{equation}
  }
}

\noindent
We show that this does indeed provide a good fit, comparable in accuracy to $\FE(a,x)$ for $\RE(a,k,p)$.  In this case, our numerical reference computation is the \fastnuf{} integration of Ref.~\cite{Upadhye:2025glk}.  As described in that reference, we account for scale-dependent growth due to neutrinos by iteratively improving the neutrino component of $\dm$ using the previous time integration, a procedure which they showed to agree with \classcode{} at the percent level.

\begin{figure}[tb]
  \includegraphics[width=76mm]{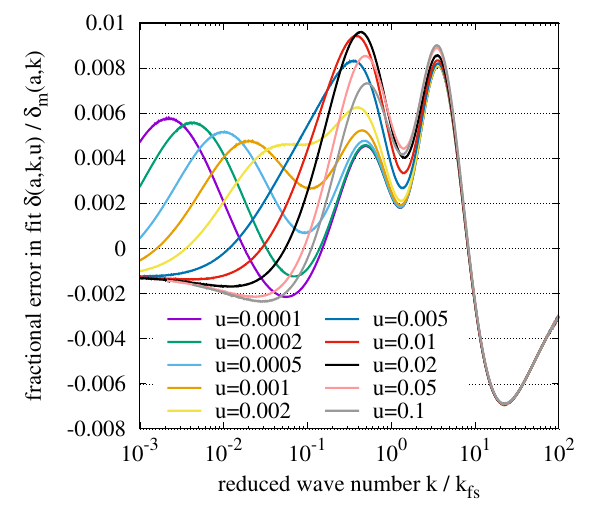}
  \includegraphics[width=76mm]{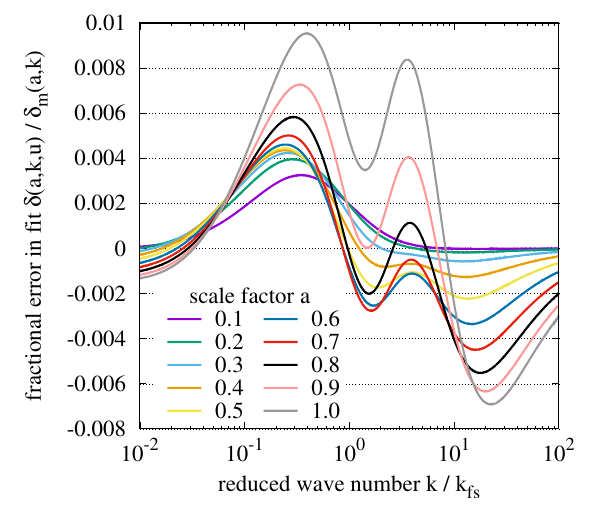}%
  
  \includegraphics[width=76mm]{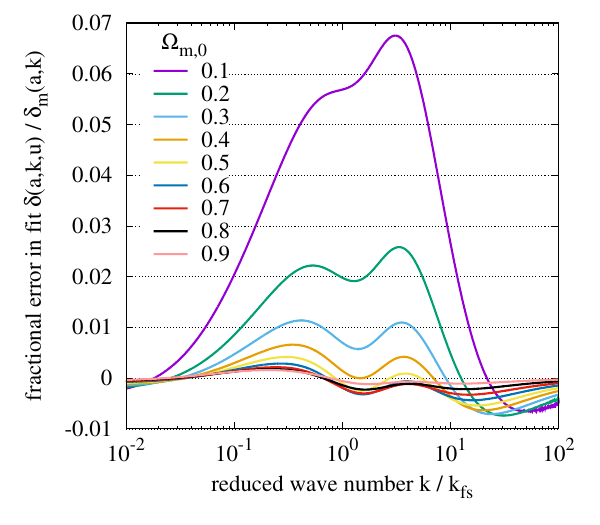}
  \includegraphics[width=76mm]{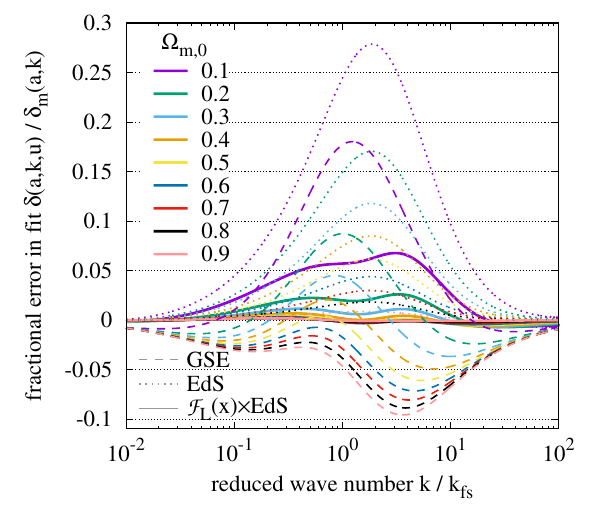}%
  
  \includegraphics[width=76mm]{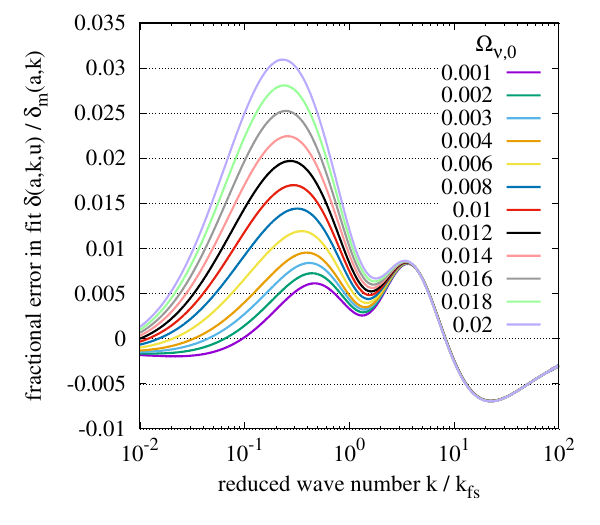}
  \includegraphics[width=76mm]{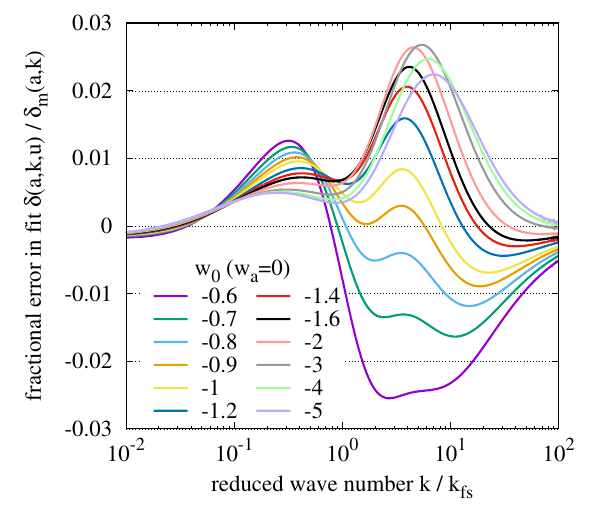}%
  \caption{
    Fractional error in the \SEEdS{} fit, \EQ{e:fit_lag} times the 
    exact EdS solution of \EQ{e:dhdm_dm_eds_lag}.  The Top-Left plot
    varies $u$ at $a=1$, while the Top-Right plot varies $a$ at 
    $u=0.01$.  The remainder of the plots
    fix $u=0.01$ and $a=1$, while varying one parameter at a time
    from the model of \EQ{e:fid_cosmo}.
    \label{f:lag_vary_params}
  }
\end{figure}

Figure~\ref{f:lag_vary_params} varies $u$ and $a$ within the fiducial model of \EQ{e:fid_cosmo}, then fixes $u=0.01$ and $a=1$ while varying $\Omo$, $\Ono$, and $w_0$ from that model, one at a time.  For the fiducial model, our fit of \EQ{e:fit_lag} is consistently accurate at the sub-percent-level.  Considering all models shown, only for $\Omo=0.1$ or $\Ono=0.02$ does the error in our fit exceed $3\%$.  Thus we demonstrate $<3\%$ accuracy for all $a$, $k$, and $u$, as well as $0.2 \leq \Omo \leq 1$, $\Ono \leq 0.018$, and $w_0 \leq -0.6$.  This is similar to the parameter range in the Eulerian case.  In particular, for N-body simulations which treat neutrinos purely non-relativistically, our fit provides a $<3\%$-accurate approximation to linear neutrinos, for a wide range of cosmological parameters, over the entirety of the simulated time period $0.01 \lesssim a \leq 1$.

The plot which looks most different from its Eulerian counterpart is Fig.~\ref{f:lag_vary_params}~(Bottom-Left), which shows more $\Ono$-dependent errors than Fig.~\ref{f:eul_vary_Onu0_w0}~(Left).  We find that this increased $\Ono$-dependence is due to the iterative procedure used by \fastnuf{} to quantify the neutrino-induced scale-dependence in the linear growth factor $D(a,k)$.  Switching off the iteration yields smaller $< 2\%$ errors which are less $\Ono$-dependent.  However, iteration was shown in Ref.~\cite{Upadhye:2025glk} to match \classcode{} results, so we regard  Fig.~\ref{f:lag_vary_params}~(Bottom-Left) as a more accurate measure of the error and its $\Ono$-dependence.

\begin{figure}[tb]
  \begin{center}
    \includegraphics[width=140mm]{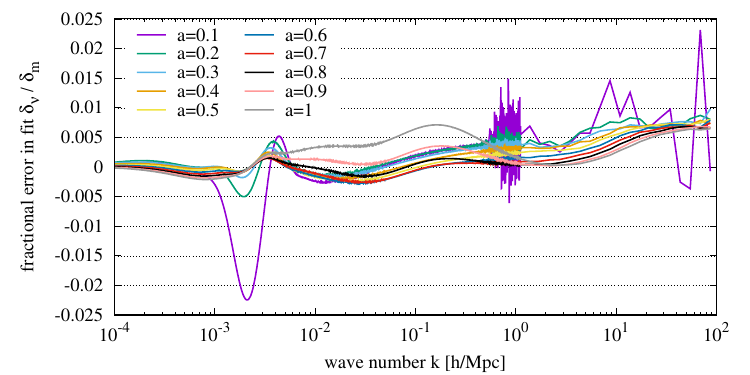}%
  \end{center}
  \caption{
    Comparison of the total neutrino density contrast, integrated over
    Lagrangian momenta and divided by $\dm$, to that
    found using \classcode{} for the fiducial model of \EQ{e:fid_cosmo}.
    \label{f:lag_dnu_dm_vs_class}
  }
\end{figure}

Finally, we compare the total neutrino density contrast for our fiducial model to \classcode{} in Figure~\ref{f:lag_dnu_dm_vs_class}.  We integrate $\delta(a,k,u)$ over Lagrangian momenta analogously to \EQ{e:dnu_tot_eul}:

\noindent
\framebox[0.99\width][c]{
  \parbox[c]{153mm}{
    \begin{equation}
      \frac{\dnu(a,k)}{\dm(a,k)}
      =
      \frac{2}{3\zeta(3)} \Fnr(a,k)
      \int_0^\infty dq \frac{q^2}{e^q+1} \FL(C(a)q)
      \left[ \frac{\dnu(a,k,u = q\Tnuo /m_\nu)}{\dm(a,k)} \right]_{\rm EdS}~.
      \label{e:dnu_tot_lag}
    \end{equation}
  }
}

\noindent
Here, $\Fnr(a,k)$, $C(a)$, $\FL(a,x)$, and $[\dnu/\dm]_{\rm EdS}$ are given by Eqs.~(\ref{e:fit_r-nr},\,\ref{e:def_GE_Ca},\,\ref{e:fit_lag},\,\ref{e:dhdm_dm_eds_lag}), respectively.  This and \EQ{e:dnu_tot_eul} are the main results of this Article.  Our public implementation\footnote{{available online at \tt{https://codeberg.org/upadhye/SE-EdS}}} uses Gauss-Laguerre quadrature, as described in Ref.~\cite{Upadhye:2024ypg}.  It also replaces the Fermi-Dirac distribution $1/(e^q+1)$ by the effective distribution ${\bar f}_\ehdm(q)$ of \EQ{e:f_edhm} and the normalization $2/(3\zeta(3))$ by $1/\int dq \, q^2 {\bar f}_\ehdm(q)$.  Once again, we apply the relativistic-neutrino correction factor of \EQ{e:fit_r-nr}.  The results are nearly identical to those of Fig.~\ref{f:eul_dnu_dm_vs_class}.  Our fit is consistently accurate to $\leq 1\%$ except for the early-time residual error $\propto (\anu/a)^2$ in \EQ{e:fit_r-nr}.  As above, the noise at $k \gtrsim 1~h/$Mpc is due to the \classcode{} neutrino density contrast.

To summarize, in this Section so far we have found simple fitting functions for the neutrino density contrasts, \EQEQ{e:fit_eul}{e:fit_lag} for Eulerian and Lagrangian momenta as well as \EQEQ{e:dnu_tot_eul}{e:dnu_tot_lag} for the total density contrasts.  We have confirmed their accuracy to $1\%-1.5\%$ for our fiducial cosmology, \EQ{e:fid_cosmo}, and to $2\%-3\%$ for $0.2 \leq \Omo \leq 1$, $0 \leq \Ono \leq 0.01$, and $w_0 \leq -0.6$. This applies to all neutrino momenta and wave numbers, to all $a \gtrsim 0.01$ in our non-relativistic-neutrino approximation and to $a \geq 0.1$ when compared to codes which account for neutrinos' early-time relativistic behavior.

\subsection{Painting with neutrinos: A simple guide}
\label{sub:fit:paint}

An important application of our fits is painting neutrinos into N-body simulations in Fourier space.  We conclude this Section by providing a guide to the painting procedure for the simple case of non-relativistic, degenerate-mass neutrinos with a mass $m_\nu$, a temperature constant $\Tnuo = 0.00016825$~eV, and a Fermi-Dirac momentum distribution.  In order to keep this discussion somewhat self-contained, we repeat essential equations from above, in addition to referencing earlier results.

Suppose that an N-body simulation employs particles only to represent the CDM and baryons, which we treat as a single fluid labeled cb.  At a given scale factor $a$, standard methods may be used to derive from the instantaneous distribution of particles a Fourier-space density contrast $\dcb(a,\vec k)$.  In a model without massive neutrinos or other HDM, this cb fluid would be the only clustering fluid, meaning that the gravitational potential $\Phi(a,\vec k)$ would be determined entirely by the distribution of particles, that is, by $\dcb(a,\vec k)$.  In turn, the gradient of $\Phi(a,\vec k)$ would determine the momentum evolution of each particle, hence its position at the next time step.

Instead, we consider a cosmology which does include massive neutrinos, of matter fraction $\fnu = \Ono / \Omo$, but we continue to use particles only for the cb fluid, of matter fraction $\fcb = 1 - \fnu$.  Painting is a procedure for estimating the total neutrino density $\dnu(a,\vec k)$ at that same time step, allowing it to be included in Poisson's equation, hence to affect the dynamics of the particles.  Our goal is thus to obtain $\dnu(a,\vec k)$ from $\dcb(a,\vec k)$.

Our first step is to find the neutrino-to-matter ratio $R$.  Applying Gauss-Laguerre quadrature to the integral in \EQ{e:dnu_tot_lag}, we obtain:
\begin{eqnarray}
R(a,k) 
  = \frac{\dnu(a,k)}{\dm(a,k)}
  &=& 
  \frac{2}{3\zeta(3)} \sum_\alpha 
  \frac{q_\alpha^2 \exp(q_\alpha) W_\alpha}{\exp(q_\alpha)+1}
  \FL(a,C(a)q_\alpha) \Reds(C(a)q_\alpha)
  \label{e:R_SEEdS_painted}
\\
C(a) &=& \sqrt{\frac{2}{3\Omo}} \frac{k \Tnuo}{\Hco m_\nu} a^{-1/2}
\nonumber\\
\FL(a,x) &=& \frac{3+x^2}{3+\sqrt{3}[1-G(a)]x + x^2}
\nonumber\\
\Reds(x) 
  &=&
  1 + 6x^2 \cos(x\sqrt{6}) \Ci(x\sqrt{6})
  + 6x^2 \sin(x\sqrt{6}) \left[\Si(x\sqrt{6}) -\frac{\pi}{2}\right].
  \nonumber
\end{eqnarray}
where $C(a)$, $\FL(a,x)$, and $\Reds(x)$ were originally given in Eqs.~(\ref{e:def_GE_Ca},\ref{e:fit_lag},\ref{e:dhdm_dm_eds_lag}), respectively.  The low-$k$ limit of the reduced growth factor, $G(a) = D(a)/a$, is given for spatially-flat models with constant dark energy equations of state in \EQ{e:G_w0cdm}.  It may be computed for more general models using standard methods, as discussed in Sec.~\ref{sub:bkg:homog}.  The $q_\alpha$ and $W_\alpha$ are, respectively, the points and weights of Gauss-Laguerre quadrature.  They may be obtained from the python {\tt scipy} package using the function {\tt special.roots\_laguerre} whose argument is the quadrature order; we find orders $\gtrsim 15$ to be adequate.

Next, we use $R(a,k)$ to obtain the painted neutrino density contrast.  In our cosmology, whose only clustering species are the cb and neutrino fluids, $\dm = \fcb \dcb + \fnu \dnu$, so we have
\begin{equation}
\dnu(a,\vec k) = 
\frac{\fcb R(a,|\vec k|)}{1-\fnu R(a,|\vec k|)}
\dcb(a,\vec k).
\label{e:dnu_SEEdS_painted}
\end{equation}
Since our \SEEdS{} fit neglects any phase differences between neutrinos and other matter, the cb and neutrino fluids differ only in their clustering amplitudes for a given Fourier mode.  

Finally, if we are uninterested in the neutrinos themselves, but only in their effect upon the gravitational potential, then we may substitute \EQ{e:dnu_SEEdS_painted} into Poisson's equation:
\begin{equation}
\Phi(a,\vec k) = 
-\frac{3 \Om(a) \Hc(a)^2}{2 |\vec k|^2}
\left[\fcb + \frac{\fnu\fcb R(a,|\vec k|)}{1-\fnu R(a,|\vec k|)}\right] \dcb(a,\vec k) .
\label{e:Phi_SEEdS_painted}
\end{equation}
In the limit that massless neutrinos become negligible, the quantity in square brackets reduces to unity, leaving Poisson's equation for the cb fluid alone.  Thus neutrinos may be painted into the simulation by including the factor in square brackets, with $R$ obtained from \EQ{e:R_SEEdS_painted}, in Poisson's equation.  At small $k = |\vec k|$, $R$ approaches unity, and the factor in square brackets again reduces to one, while at large $k$, $R$ becomes small, and the factor approaches $\fcb$.  The key quantity in both \EQEQ{e:dnu_SEEdS_painted}{e:Phi_SEEdS_painted}, $R(a,k)$, is computed by our public code ({\tt codeberg.org/upadhye/SE-EdS}) for $\Omo$, $w_0$, and neutrino masses chosen by the user.

\section{Generalizations}
\label{sec:gen}

Our \SEEdS{} fitting functions, \EQEQ{e:fit_eul}{e:fit_lag}, depend upon the cosmological parameters only indirectly, through $G(a)$, so they are potentially much more general than the $w_0$CDM models  considered in previous sections.  Section~\ref{sub:gen:OKw0wa} verifies that the fits are accurate in cosmologies with multiple HDM species, spatial curvature, and dark energy with a time-dependent equation of state.  In Sec.~\ref{sub:gen:cdm-de_coupling} we confirm the accuracy of \SEEdS{} for a simple phenomenological approximation to models with couplings between dark matter and dark energy, which have been proposed as solutions to the $\Hco$ and $\sigma_8$ tensions. 

A major motivation for this Article is the possibility of a simple fitting function with which to paint HDM into N-body simulations in Fourier space, yielding a new linear response method.  Section~\ref{sub:gen:nl_cdm} compares neutrino linear response using \SEEdS{} and GSE, finding \SEEdS{} to be more accurate in the linear neutrino regime even at late times.  At early times, \SEEdS{} is highly accurate by design, as the cosmology approaches EdS, making our fit a fast, accurate, and inexpensive method for quantifying the effects of scale-dependent HDM clustering upon other species.

\subsection{Masses, curvature and evolving dark energy}
\label{sub:gen:OKw0wa}

 We begin by studying three important generalizations, multiple HDM masses, spatial curvature, and a time-variation in the dark energy equation of state.  We treat curvature and dark energy within the formalism of  Ref.~\cite{Linder:2003dr} as a non-clustering ``dark energy'' component as discussed in Sec.~\ref{sub:bkg:homog}.  In this formalism, curvature corresponds to a constant equation of state $w_K = -1/3$ and a density fraction today of $\OKo$.

\begin{figure}[tb]
  \includegraphics[width=75mm]{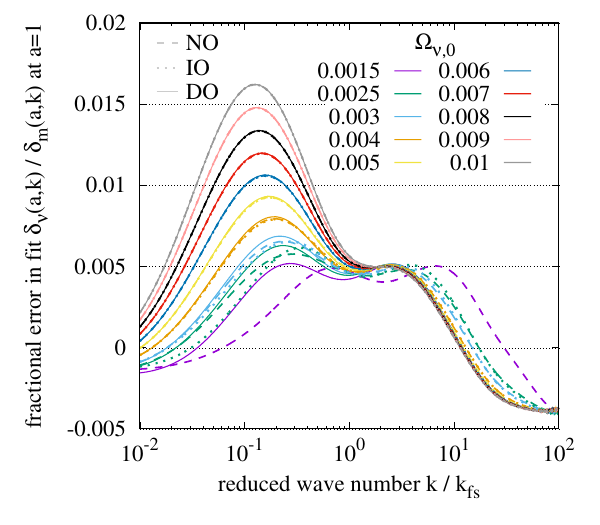}
  ~\includegraphics[width=75mm]{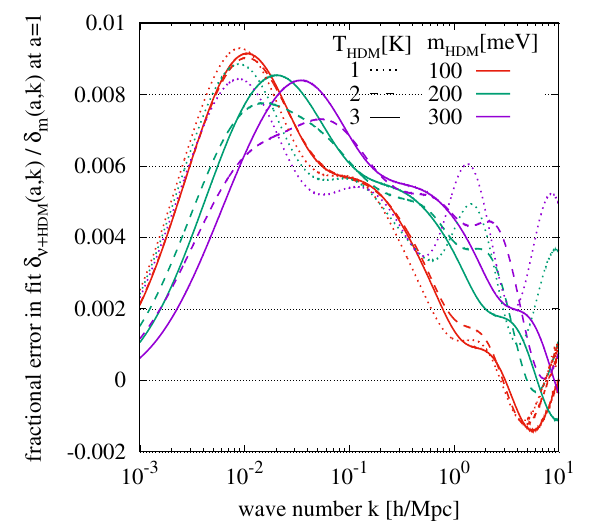}%
  \caption{
    Tests of our fitting function \EQ{e:fit_lag} integrated over momenta,
    for two different modifications to the HDM sector, compared to the
    \fastnuf{} calculation of Ref.~\cite{Upadhye:2025glk}.
    (Left)~Neutrinos with normal (NO) and inverted (IO) mass orderings
    are compared to the degenerate (DO) ordering considered thus far.
    (Right)~A single Bosonic HDM particle of mass $m_\hdm$ and temperature
    $T_\hdm$ is added to our fiducial model, \EQ{e:fid_cosmo}, for which
    $\Mnu = 162$~meV and $\Tnuo = 1.95$~K.
    \label{f:lag_vary_ord_hdm}
  }
\end{figure}

Until now, we have considered three degenerate-mass neutrinos with no other HDM species.  Neutrinos of different masses consistent with measured mass splittings, as well as additional HDM particles such as axions, may be included through the EHDM formalism of Sec.~\ref{sub:bkg:ehdm}, which reweights the same set of Lagrangian flows to represent different HDM species.  Figure~\ref{f:lag_vary_ord_hdm} considers two such variations.  The left plot confirms that replacing DO neutrinos with NO or IO neutrinos does not appreciably affect the error in the \SEEdS{} fits.  This is not surprising, as the smallest and largest neutrino masses only differ substantially for $\Ono \lesssim 0.005$, where \SEEdS{} errors are already $\lesssim 1\%$.  Meanwhile, Fig.~\ref{f:lag_vary_ord_hdm}~(Right) adds a single thermalized Bosonic HDM particle, described by a Bose-Einstein distribution, to our fiducial model.  The maximum error in the \SEEdS{} fit for the nine models considered is $< 1\%$, not much larger than the maximum error at $a=1$ in Fig.~\ref{f:lag_dnu_dm_vs_class}.  Thus we have confirmed the accuracy of our fit for different HDM masses, temperatures, and distribution functions.

\begin{figure}[tb]
  \includegraphics[width=75mm]{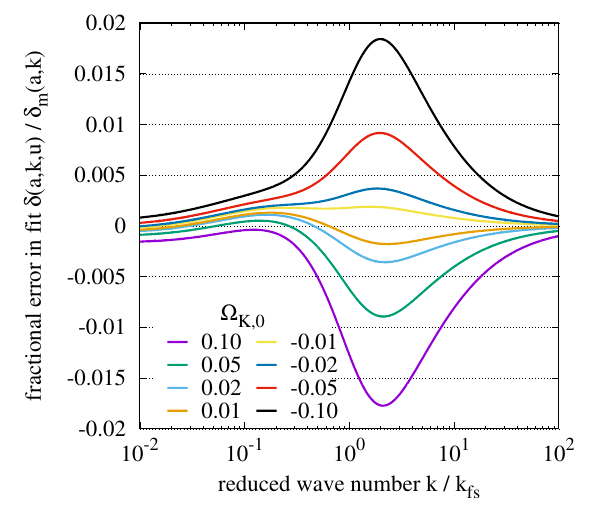}
  ~\includegraphics[width=75mm]{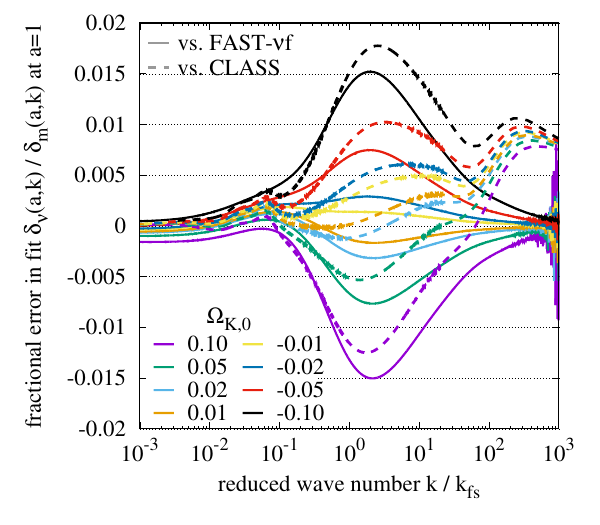}%
  \caption{
    Tests of our fitting function \EQ{e:fit_lag} as  $\OKo$ is varied, at $a=1$,
    in models containing only clustering matter and curvature.  Aside from
    $w_0=-1/3$ and $\Omo + \OKo = 1$, other parameters are set by
    \EQ{e:fid_cosmo}. 
    (Left)~Comparison to \fastnuf{} (Ref.~\cite{Upadhye:2025glk}) for
    Lagrangian velocity $u=0.01$.
    (Right)~Comparison of the total neutrino density contrast to \fastnuf{}
    (solid) and \classcode{} (dashed).
    \label{f:lag_vary_OK}
  }
\end{figure}

Next, consider curvature. Extrapolating Fig.~\ref{f:lag_vary_params}~(Bottom-Right) suggests that $w_K=-1/3$ will have an error well in excess of our $3\%$ error threshold.  However, given our fiducial model, this would imply an $\OKo$ of $0.67$, some two orders of magnitude larger than the constraints of Ref.~\cite{Planck:2018vyg}.  We restrict ourselves here to $|\OKo| \leq 0.1$.  The simple case of models containing only matter and a curvature can already be handled in our existing $w_0$CDM formalism  by setting $w_0=-1/3$.  Figure~\ref{f:lag_vary_OK} tests our fit for the Lagrangian-momentum case for this set of models and confirms its accuracy to $<2\%$ for the entire range  $|\OKo|\leq 0.1$.

In Fig.~\ref{f:lag_vary_OK}~(Right), two different high-$x$ errors are evident.  The \fastnuf{} method uses $\Si(y)$ and $\Ci(y)$, which are highly oscillatory at small scales, leading to a sub-percent-level noise at $x = 1000$ which rapidly diminishes at lower $x$.  Meanwhile, the \classcode{} comparison has a percent-level offset at $x \gtrsim 100$, where we expect close agreement with the free-streaming limit.  This is likely a numerical error, as it is worsened by increasing the \classcode{} error tolerances.  Since Sec.~\ref{sec:fit} and Fig.~\ref{f:lag_vary_OK} all demonstrate the close agreement of our Eulerian and Lagrangian-momentum fits at $a=1$ with \classcode{}, \fastnuf{}, and a direct numerical integration of $\psi$ in \EQ{e:psi}, for a wide range of models,  we limit ourselves to comparisons with \fastnuf{}, which is the fastest, and to the Lagrangian-momentum-fit, for the remainder of Sec.~\ref{sub:gen:OKw0wa}. 

Finally, we consider a universe containing: matter, of density fraction $\Omo$; curvature, of fraction $\OKo$ and  equation of state $w_K=-1/3$; and Chevallier-Polarski-Linder (CPL) dark energy, with fraction $\Odo$ and equation of state parameterized by $w(a) = w_0 + w_a(1-a)$ as described in Refs.~\cite{Chevallier:2000qy,Linder:2002et}.  For this cosmology,  $\wLJ(a)$ and $\XLJ(a)$ of \EQ{e:wLJ_XLJ} become:
\begin{eqnarray}
  \wLJ(a) &=&
  \frac{w_K\OKo a^{-3w_K} + (w_0+w_a(1-a))\Odo a^{-3w_0-3w_a} e^{-3w_a(1-a)}
  }{
    \OKo a^{-3w_K} + \Odo a^{-3w_0-3w_a} e^{-3w_a(1-a)}
  },
  \\
  \XLJ(a) &=&
  \frac{\Omo}{\OKo a^{-3w_K} + \Odo a^{-3w_0-3w_a} e^{-3w_a(1-a)}}.
\end{eqnarray}
We integrate \EQ{e:GLJ} for $G(a)$, which we apply to our fitting function \EQ{e:fit_lag}. 

\begin{figure}[tb]
  \includegraphics[width=75mm]{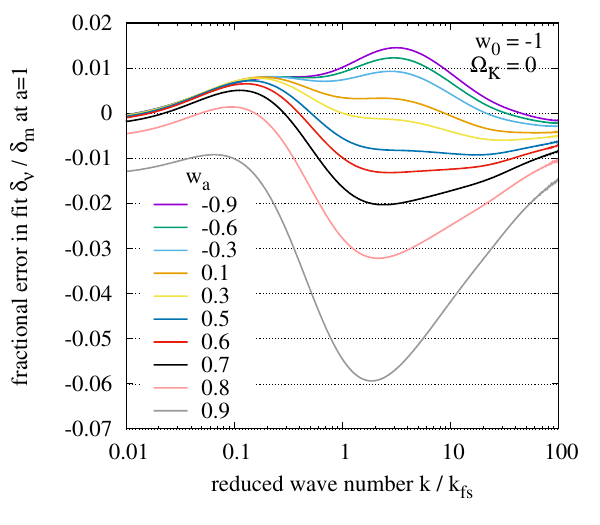}
  ~\includegraphics[width=75mm]{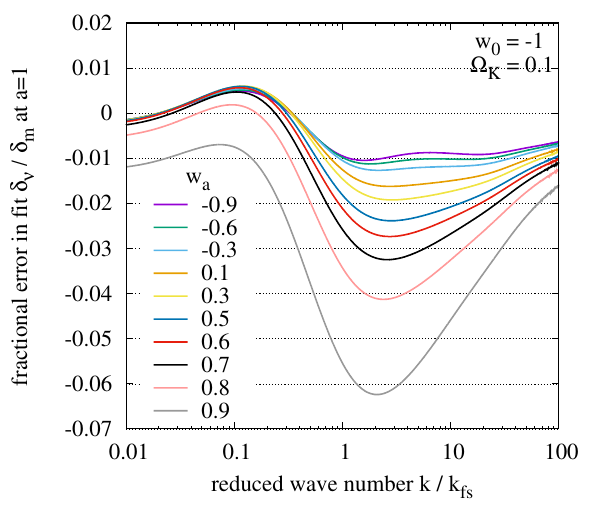}%

  \includegraphics[width=75mm]{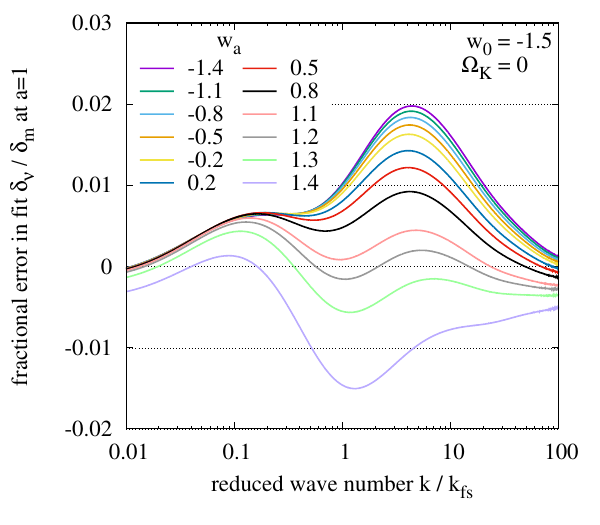}
  ~\includegraphics[width=75mm]{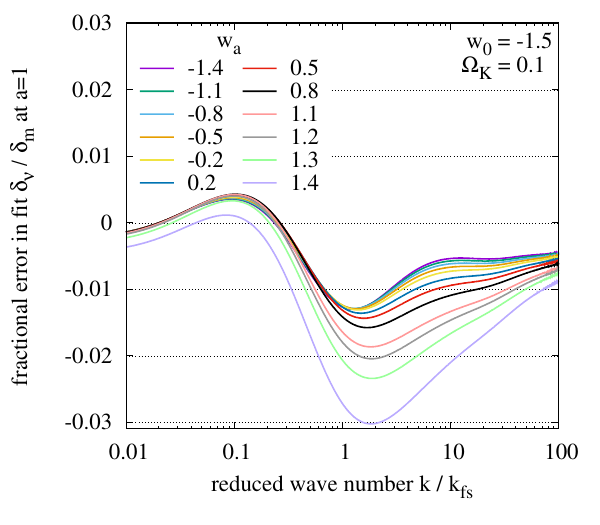}%
  \caption{
    Tests of the fitting function of \EQ{e:fit_lag} integrated over Lagrangian
    momenta, at $a=1$, as $w_0$, $w_a$, and $\OKo$ are varied. The dark energy
    density fraction is $\Odo = 1 - \Omo - \OKo$, and all other
    parameters are fixed to \EQ{e:fid_cosmo}.
    Left column:~$\OKo=0$. Right column: ~$\OKo=0.1$.
    Top row: $w_0=-1$. Bottom row: $w_0=-1.5$.
    \label{f:lag_vary_wa}
  }
\end{figure}

Figure~\ref{f:lag_vary_wa} tests our fit for a range of $w_a$ at $w_0=-1$ (Top row) and $w_0=-1.5$ (Bottom row), with $\OKo=0.0$ (Left column) and $\OKo=0.1$ (Right column) used to estimate the effects of curvature.   As $w_0+w_a$ approaches zero from below, the CPL equation of state behaves as an early dark energy, with a non-negligible dark energy fraction even at early times $a \sim 10^{-3}$.  For example, with $w_0=-1$, $w_a=0.9$, and $\OKo=0$, the dark-energy-to-matter ratio $\XLJ^{-1}$ for the model in the figure is $0.017$ at $a=10^{-3}$ and $0.035$ at $a=10^{-2}$.  Evidently, increasing $w_a$ also leads to greater errors in the fit.  This qualitatively agrees with Fig.~\ref{f:lag_vary_params}~(Bottom-Right), confirming that dark energy which represents a sizable fraction of the energy density over a large range of scale factors increases the error in our fit.  This error is exacerbated by $\OKo>0$, which also leads to early-time deviations from EdS. Reducing $w_a$ at $w_0=-1$, we find errors $<3.5\%$ for $w_a \leq 0.8$ when $\OKo=0$, and for $w_a \leq 0.7$ when $\OKo=0.1$, a significant portion of the CPL and curvature parameter space.  Reducing $w_0$ to $-1.5$, we find errors $\leq 3\%$ even for $w_a=1.4$, since the reduced late-time equation of state leads to a smaller $\XLJ^{-1}(10^{-3}) \approx 0.004$ for that model.

\begin{figure}
  \begin{center}
    \includegraphics[width=140mm]{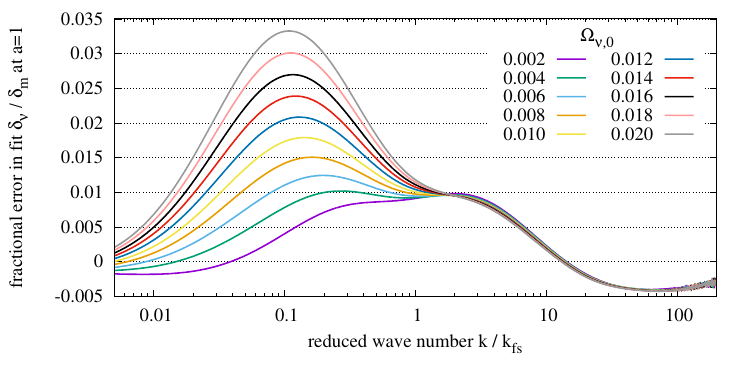}%
  \end{center}
  \caption{
    Tests of the \SEEdS{} fit of \EQ{e:fit_lag}, integrated over Lagrangian
    momenta, for the best-fit CPL dark energy of Ref.~\cite{DESI:2025ejh},
    $w_0 = -0.787$ and $w_a = -0.63$, at $a=1$.  \SEEdS{} is accurate to
    $\leq 3.0\%$ for $\Ono \leq 0.018$ and to $\leq 1.0\%$ for
    $\Ono \leq 0.004$.
    \label{f:desi_bf}
  }
\end{figure}

A timely application of our fitting function is to the best-fit evolving dark energy model of the DESI neutrino analysis of Ref.~\cite{DESI:2025ejh}.  Whereas their $\nu\Lambda$CDM parameter constraints prefer a negative sum of neutrino masses at the $68\%$ confidence level, an obviously unphysical bound, their allowing $w_0$ and $w_a$ to vary relaxes this $\Mnu$ bound sufficiently to allow positive values at $68\%$ confidence.   Figure~\ref{f:desi_bf} demonstrates the accuracy of the \SEEdS{} fit for their best-fit dark energy, $w_0 = -0.787$ and $w_a = -0.63$, for a range of neutrino fractions.  Consistent with the results of Sec.~\ref{sec:fit}, we find $\leq 3\%$ accuracy for $\Ono \leq 0.018$, and $\leq 1\%$ accuracy for $\Ono \leq 0.004$.  With a prior $\Mnu\geq 0$, the DESI analysis constrains $\Ono \leq 0.0031$ at $95\%$ confidence, a region over which the \SEEdS{} fit is accurate at the percent level.

In summary, we have demonstrated that our \SEEdS{} fitting functions of Sec.~\ref{sec:fit} are accurate not just for the linear $w_0$CDM models considered earlier, but for a much wider range of linear models with multiple HDM species, spatial curvature, and dark energy with a time-dependent equation of state.  The models for which our fit performs the worst are early dark energies, which cause the universe to deviate from the EdS case even at early times.

\subsection{CDM-dark-energy coupling}
\label{sub:gen:cdm-de_coupling}

Recent proposals for addressing the cosmological tensions, Refs.~\cite{Giare:2025ath,Yang:2026qmz,Ambelu:2026mkd,Sabogal:2026bsw}, have suggested that the CDM and dark energy could be coupled, potentially suppressing CDM clustering and changing the dark energy equation of state.  Since a full exploration of the parameter space is beyond the scope of this Article, we focus on the phenomenological approximation of Ref.~\cite{Giare:2025ath}, which leaves the late-time homogeneous background evolution unchanged while modifying the matter growth rate using a constant parameter $\gamma$:
\begin{eqnarray}
\frac{d \log(D)}{d \log(a)} &=& \Om(a)^\gamma
\label{e:CDM-DE_f}
\\
G(a) &=& 
\frac{r_0^{\frac{1}{3w_0}}}{a}
\exp\left[
    \frac{1}{3w_0} \left[
    \gamma^{-1} \Om(a)^\gamma {}_2F_1(1,\gamma,1+\gamma,\Om(a)) 
    + \psi(\gamma) - \psi(1)
    \right]
\right].
\label{e:CDM-DE_G}
\end{eqnarray}
Here, $\psi(z) := d \log(\Gamma(z))/dz$ is the digamma function, and we have assumed the constant-$w$ ($w_0$CDM) homogeneous background cosmology of \EQ{e:Oma_Oda_w0cdm}.  For $\gamma$ of $0.55\approx 5/9$, these formulae approximate the exact ones of our fiducial model of \EQ{e:fid_cosmo}.  A larger $\gamma$ will suppress the late-time growth rate, while a smaller one will increase it.  The scaling with $\Om(a)$ ensures that the effects of this modification diminish at early times as $\Od(a)$ diminishes, leading $G(a)$ of \EQ{e:CDM-DE_G} to approach unity.

\begin{figure}[tb]
\hskip-1mm\includegraphics[width=79mm]{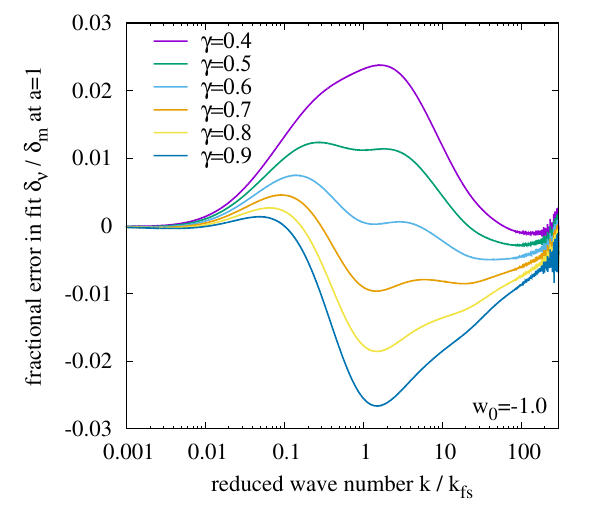}
~\includegraphics[width=79mm]{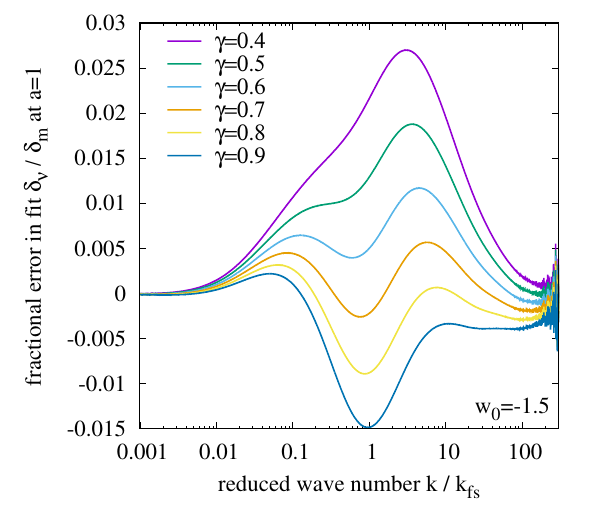}
\caption{
  Tests of the integrated \SEEdS{} fit of \EQ{e:dnu_tot_lag} for a set of 
  phenomenological CDM-DE coupling models, in which the matter growth rate
  is modified to $d\log(D)/d\log(a) = \Om(a)^\gamma$.  (Left)~$w_0=-1$,
  (Right)~$w_0=-1.5$, with other parameters from \EQ{e:fid_cosmo}. 
  \label{f:cdm-de_coupling}
} 
\end{figure}

Figure~\ref{f:cdm-de_coupling} tests the integrated \SEEdS{} fit of \EQ{e:dnu_tot_lag} for several $\gamma$ values, and for two values of the constant dark energy equation of state.  Since the $a=1$ errors shown are larger than earlier-time errors, we find the fit to be $<3\%$ accurate at all times and wave numbers for a wide range $0.4 \leq \gamma \leq 0.9$ about our fiducial cosmology.  This $\gamma$ range represents changes in the $a=1$ growth factor ranging from $+5.7\%$ to $-11\%$, which is more than sufficient to cover the range of $\sigma_8$ values preferred by recent analyses such as Refs.~\cite{Planck:2018vyg,DESI:2024hhd,Stolzner:2025htz}.  For $w_0<-1$, preferred by Ref.~\cite{Yang:2026qmz}, $\Od(a)$ diminishes more quickly as $a$ decreases, thereby reducing the error in our fit, particularly at high $\gamma$.

\subsection{Non-linear CDM and baryons}
\label{sub:gen:nl_cdm}

The GSE approximation summarized in Sec.~\ref{sub:bkg:nupt} was derived for an arbitrary gravitational potential, with the only constraint being that the HDM obeys linear equations of motion.  This made it useful as a linear response method, in which HDM evolves linearly with a non-linear $\dm$ source.  Linear response allows the CDM and baryonic contributions to $\dm$ to be drawn from non-linear perturbation theory or N-body simulations, as described in Refs.~\cite{Brandbyge:2008js,Ali-Haimoud:2012fzp}.  Section~\ref{sub:fit:paint} implemented linear response using our \SEEdS{} fit.  This is a useful approximation for those not interested in the small-scale clustering of HDM itself, but only its impact upon the clustering of CDM and baryons.

However, our EdS-test-HDM approximation of Sec.~\ref{sec:eds} explicitly invoked linearity for the CDM and baryons.  Thus the question arises: can our \SEEdS{} fit serve as a linear response method for neutrinos and other HDM?  Since non-linearities are weak at early times, \SEEdS{} will be accurate there.  Since it interpolates between the same clustering and free-streaming limits as the GSE fit, our fit should not be significantly worse even at late times.  Here we test whether it improves upon GSE at late times.

As a non-linear matter source, we use the emulated $\dm(a,k)$ of Ref.~\cite{Upadhye:2025glk}.  At early times, $a < 1/3$, it uses the Time-RG perturbation theory of Refs.~\cite{Pietroni:2008jx,Lesgourgues:2009am,Upadhye:2017hdl,Chen:2020bdf}.  At later times, its CDM and baryonic density contrast is drawn from the emulator of Ref.~\cite{Moran:2022iwe}, while its neutrino contribution comes from the non-linear perturbative training set of Refs.~\cite{Chen:2022cgw,Upadhye:2023bgx}.  With this $\dm(a,k)$ as a source, Ref.~\cite{Upadhye:2025glk} uses \fastnuf{} as a linear response method which tracks individual Lagrangian flows as well as the total linear neutrino density contrast.  Here, we use \fastnuf{} linear response as our reference, against which we test our fit of \EQ{e:fit_lag}.

\begin{figure}[tb]
  \hskip-2mm\includegraphics[width=79mm]{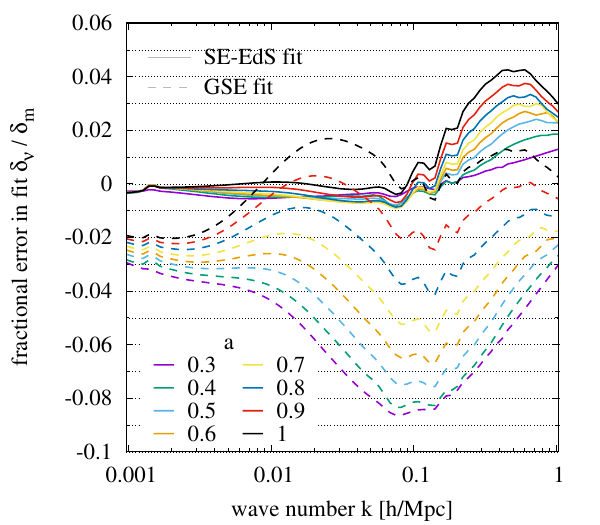}
  ~\includegraphics[width=79mm]{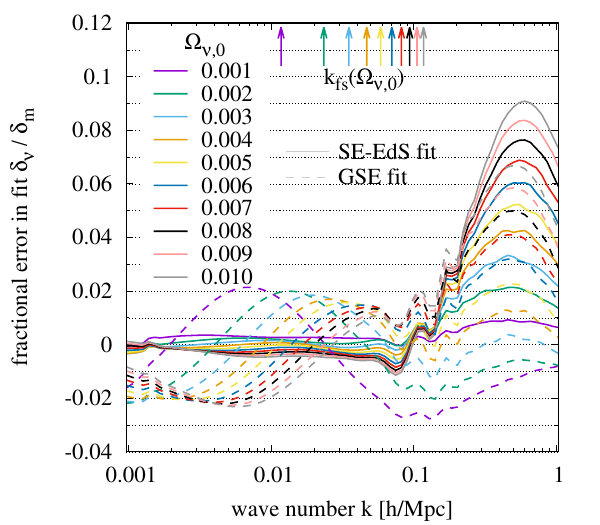}%
  \caption{
    Tests of the \SEEdS{} fit of \EQ{e:fit_lag} (solid) and the GSE fit of
    \EQ{e:gse_lag} (dashed) for linear response to a non-linear $\dm(a,k)$.
    (Left)~Variation of $a$ for the fiducial cosmology of \EQ{e:fid_cosmo}.
    (Right)~Variation of $\Ono$ at $a=1$, with all other parameters fixed to
    their fiducial values.  Arrows across the top of the plot show the
    free-streaming scale of \EQ{e:fit_se} associated with each $\Ono$.
    \label{f:dm_nL}
  }
\end{figure}

Figure~\ref{f:dm_nL} shows our results for the \SEEdS{} and GSE fits.  Consider first the Left plot.  \SEEdS{} is highly accurate at all $a$ for $k \leq 0.1~h/$Mpc, but its fractional error grows after that, rising to $4.3\%$ at $a=1$ and $k \approx 0.5~h/$Mpc.  Reducing $a$ causes the magnitude of this maximum error to fall and its location to shift to greater $k$, even though the free-streaming wave number drops with $a$.  Thus the error is associated with non-linear growth in $\dm(a,k)$.  Specifically, non-linear growth increases $\dm(a,k)$ more at high $k$ than the corresponding linear response increases $\dnu(a,k)$, causing the ratio $\dnu/\dm$ to fall, hence our fit to overestimate it.  By comparison, the GSE fit has a lower maximum error, about $2\%$, at $a=1$.  However, at low $k$, it approaches the correct $\dnu(a,k)$ only slowly, since it lacks the EdS NLO corrections identified in Sec.~\ref{sub:eds:lag}.  This behavior was noted earlier in Fig.~\ref{f:eul_vary_Om0}~(Right) and Fig.~\ref{f:lag_vary_params}~(Center-Right).  Furthermore, its errors grow as $a$ is reduced, exceeding the \SEEdS{} errors for $a \leq 0.8$.

Since \SEEdS{} errors are greatest at $a=1$, Fig.~\ref{f:dm_nL}~(Right) fixes this value and varies $\Ono$.  The maximum error is $\approx 9\%$ for $\Ono=0.01$, and scales approximately as $\kfs / (1~h/{\rm Mpc})$.  Once again, we see that the location of the maximum error is uncorrelated with the free-streaming scale.  Furthermore, at $\Ono \gtrsim 0.005$, GSE also overpredicts the high-$k$ numerical clustering.  In both fitting functions, this error is falling at $k=1~h/$Mpc, as we expect from the fact that the free-streaming limits of Sec.~\ref{sub:bkg:nupt} do not assume a linear $\dm(a,k)$.

Of course, linear response itself underpredicts the small-scale neutrino density due to neglecting non-linear corrections to $\dnu(a,k)$.  Using the emulator of Ref.~\cite{Upadhye:2025glk} to estimate these fractional non-linear corrections for the $\Ono=0.010$ model of Fig.~\ref{f:dm_nL}~(Right), we find non-linear enhancements to $\dnu(1,k)$ of $4\%$, $16\%$, $39\%$, and $64\%$ at $k$ of $0.1~h/$Mpc, $0.2~h/$Mpc, $0.5~h/$Mpc, and $1~h/$Mpc, respectively.  Thus the \SEEdS{} fit overestimates linear response precisely where linear response more significantly underestimates the non-linear $\dnu$.  The high-$k$ overestimate of Fig.~\ref{f:dm_nL} is therefore a subdominant error.  Meanwhile, in the $k \lesssim 0.1~h/$Mpc where linear response is highly accurate, \SEEdS{} is as well, with a maximum error $\approx 1\%$ at $a=1$ over the range $\Ono \leq 0.01$ covered by the plot.

For applications insensitive to this subdominant high-$k$ error, \SEEdS{} appears to be a more accurate neutrino clustering approximation than GSE.  Firstly, it has smaller errors in the $k \lesssim \kfs(a)$ regime where neutrinos will have the greatest impact on CDM and baryonic clustering.  Secondly, \SEEdS{} is significantly more accurate at early times.  N-body simulations typically integrate over the range $0.01 \lesssim a \leq 1$.  Judging from Fig.~\ref{f:dm_nL}~(Left), such a simulation using GSE to approximate $\dnu$ would underestimate the early-time, large-scale neutrino contribution to the gravitational potential by $\sim 10\%$, leading to a slight underprediction of the impact of neutrinos upon the CDM and baryonic clustering.  \SEEdS{}, by contrast, is designed to approach EdS as $G(a) \rightarrow 1$, ensuring accuracy in the matter-dominated era.

\section{Correspondence with higher-order corrections}
\label{sec:nlo}

An alternative approach to improving upon the GSE fits of Sec.~\ref{sub:bkg:nupt} is to begin with the integral solutions of \EQEQ{e:deltaf_eul_lin_ell0}{e:delta_lag_lin_ell0} and derive next-to-leading-order terms in both the clustering and free-streaming limits.  However, as noted at the end of Sec.~\ref{sec:eds}, the exact EdS solution has non-perturbative terms $\sim x^2 \log(x)$ in the clustering limit.  We do not find any perturbative low-$x$ expansion which improves upon the GSE fit at $x \approx 0.1$.

In the free-streaming limit, on the other hand, the exact EdS solutions have asymptotic approximations, polynomials in $x^{-1}$, with no non-perturbative terms.  Including more terms in the asymptotic expansions improves agreement with the EdS solutions at $x \gtrsim 10$.  The difficulty is that the coefficients of the $x^{-n}$ terms rise rapidly with $n$.  Thus, beyond some $n$, including more higher-order terms worsens the fit in the $x \lesssim 10$ range where errors in the GSE fit are largest.  In the Lagrangian-momentum case, the NLO free-streaming correction to $\delta(a,k,u)/\dm(a,k)$ is proportional to $x^{-4}$, two powers of $x^{-1}$ greater than the leading-order term $x^{-2}$.  By contrast, the NLO terms in both the GSE and \SEEdS{} fits are $\propto x^{-3}$.  We find that even the NLO Lagrangian-momentum free-streaming correction worsens the fit.

The NLO free-streaming Eulerian correction, on the other hand, is of order $x^{-3}$.  Remarkably, for $w_0=-1$ and low $a$, it matches the $x^{-3}$ term of the \SEEdS{} fit to a few percent.  Thus it automatically satisfies the first two EdS limits of Sec.~\ref{sub:fit:3limits}, and its accuracy is only slightly worse than that of \SEEdS{}.  However, it fails to satisfy the third EdS condition, since its perturbative approximation $|\partial^n \Phi/\partial s^n| \sim \Hc^n |\Phi|$ (with $\Hc \ll k$) fails for large $|w_0|$.  We can see this by noting that $\Phi \propto G(a)$ and using \EQ{e:G_large_w0}, whose time-dependence arises entirely from $\Om(a)$ of \EQ{e:Oma_Oda_w0cdm}.  The leading-order term in $d^n \Om(a) / ds^n$ is $\propto |w_0 \Hco|^n$ at large $|w_0|$.  Thus \SEEdS{} may be considered a generalization of this NLO GSE fit to a broader range of dark energies.  In Sec.~\ref{sub:nlo:eul} we derive this higher-order correction in the Eulerian case and demonstrate its agreement with \SEEdS{} for $w_0=-1$, while in Sec.~\ref{sub:nlo:lag} we find the  expansion in the Lagrangian-momentum case and show that it lacks an $x^{-3}$ term.

\subsection{Eulerian free-streaming expansion}
\label{sub:nlo:eul}

Motivated by \EQ{e:eom_eul}, we define $\IEc_0(k,y)$ as:
\begin{equation}
  \IEc_0(k,y) := \int_y^{\yin} dy' a(y')^2 \Phi(y',k) W(y'-y)
  \quad\Rightarrow\quad
  \dfmu = \frac{m_\nu^2}{p^2} \dfdlnp \IEc_0(k,y).
\end{equation}
Defining $\UEc_0(y') = a(y')^2 \Phi(y')$ (with the $k$-dependence suppressed for clarity) and $\VEc_0(y') = -\sinc(y'-y) = \sin(y'-y)/(y'-y)$, we see that $\IEc_0 = \int \UEc_0 d\VEc_0$.  Integrating by parts, we find that $\IEc_0 = \left.[\UEc_0(y') \VEc_0(y')]\right|_y^{\yin} - \int \VEc_0 d\UEc_0$.  This is dominated by the first term, the boundary term, which in the limit $\yin\rightarrow\infty$ approaches $a(y)^2 \Phi(y)$.  This is precisely the argument already made in \EQEQ{e:eul_int_by_parts}{e:def_kfs_eul}.

The NLO term arises from $\int \VEc_0 d\UEc_0$.  Integrating by parts again, we find that
\begin{eqnarray}
  \IEc_1(k,y) &:=& -\int_y^{\yin} dy' \VEc_0 \frac{d\UEc_0}{dy'}
  = \int_y^{\yin} dy' \sinc(y'-y) a(y')^2 \Phi(y',k)
  = \int \UEc_1 d\VEc_1
  \\
  \UEc_1(y') &=& \frac{\partial}{\partial y'} (a(y')^2 \Phi(y',k)),
  \quad
  \VEc_1(y') = \si(y'-y) = \Si(y'-y) - \frac{\pi}{2}
  \nonumber\\
  \Rightarrow
  \IEc_1 &=&
  \left.[\UEc_1(y')\VEc_1(y')]\right|_y^{\yin} - \int \VEc_1 d\UEc_1
  \rightarrow
  \frac{\pi}{2} \frac{\partial}{\partial y} [a(y)^2 \Phi(y,k)]
  - \int \VEc_1 d\UEc_1~\quad
\end{eqnarray}
where the limit in the final line is $\yin \rightarrow \infty$.  Once again, the boundary term is the dominant one.  Thus our NLO approximation to $\IEc_0$ and $\dfmu$ is
\begin{equation}
  \IEc_0(k,y) \approx a^2 \Phi
  + \frac{\pi}{2} \frac{\partial (a^2 \Phi)}{\partial y} 
  ~\Rightarrow~
  \dfmu
  \approx \frac{m_\nu^2}{p^2} \dfdlnp \left[
    a(y)^2 \Phi(y) +
    \frac{\pi}{2} \frac{\partial}{\partial y} (a(y)^2 \Phi(y,k))
    \right]
  \label{e:eul_nlo}
\end{equation}

We could continue integrating by parts, defining $\IEc_2 = -\int \VEc_1 d\UEc_1 = \int \UEc_2 d\VEc_2$ with $\UEc_2 = \partial^2(a^2\Phi)/\partial (y')^2$ and $\VEc_2 = (y'-y)\si(y'-y) + \cos(y'-y)$.  However, let us stop here, noting that higher-order terms involve more derivatives of the gravitational potential, $\UEc_n\VEc_n \sim \partial^n (a^2\Phi) / \partial y^n$.  In linear EdS, where $\Phi$ is constant and $a(y) = [2kp/(\Hco m_\nu y)]^2$,
\begin{equation}
  \frac{d^n}{dy^n}[a(y)^2 \Phi]
  = \frac{1}{6} (-1)^n(n+3)!\left(\frac{m_\nu \Hc(a) a}{2pk}\right)^n a^2 \Phi 
  \qquad\mathrm{(linear~EdS)}.
  \label{e:da2Phi_dy_EdS_lin}
\end{equation}
Since $m_\nu a \Hc / p \propto \kfs(a,p)$, the $n$-th-order term is proportional to $(\kfs/k)^n = x^{-n}$.  Thus it is precisely in the free-streaming limit that this series is expected to converge.  This property that higher-order corrections are suppressed by more powers of $x$ is also true of $\Lambda$CDM and similar models, but, as we pointed out at the beginning of this Section, higher orders in large-$|w_0|$ models also contain higher powers of $|w_0|$, spoiling convergence.

Consider the NLO term of \EQ{e:eul_nlo}.  In linear theory, Poisson's equation implies that $\Phi \propto G = D/a$.  The first derivative of $a^2 G$ may be found exactly for $w_0$CDM models using \EQEQ{e:G_w0cdm}{e:dDda_w0cdm}.  From \EQ{e:eul_nlo} we may compute $\dfmu/\dm$ for $w_0$CDM as well as EdS.  Their ratio is an approximation to $\RE(a,k,p)$ of \EQ{e:RE_RL} at high $k$, whose leading-order term is $1$ and whose NLO term is $\propto x^{-1}$.  The next step is to find a \SEEdS{}-like rational function fit which reproduces the NLO expansion at high $x$ as well as $\RE \rightarrow 1$ at low $x$.  We find
\begin{equation}
  \FNLO(a,x) = \frac{9+x^2}{9+3g(a)x + x^2}
  \quad\mathrm{where}\quad
  g(a) = \frac{\pi}{3}\sqrt{\frac{2}{3}}
  \left(
  \frac{G(a)+\tfrac{dD}{da}}{2G(a) \Om(a)^{1/2}} - 1
  \right).
  \label{e:fit_nlo}
\end{equation}
Note that the quantity in large parenthesis in the expression for $g$ vanishes when $\Omo$ approaches unity, as well as when $a \rightarrow 0^+$ in $w_0$CDM models.  Thus $\FNLO$ automatically satisfies the first two EdS limits of Sec.~\ref{sub:fit:3limits}.

\begin{figure}[tb]
  \begin{center}
    \includegraphics[width=140mm]{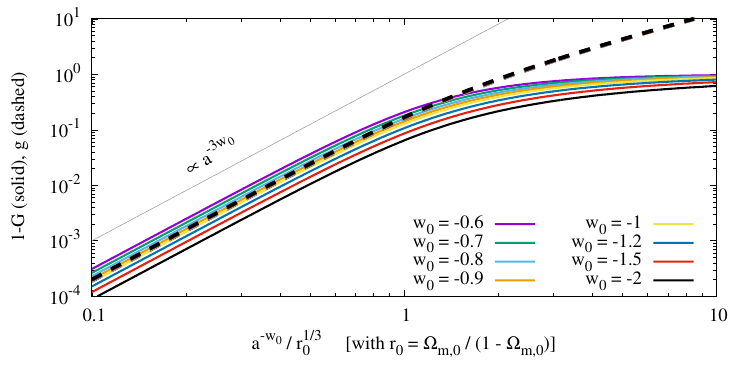}%
  \end{center}
  \caption{
    Comparison between $1-G(a)$ (solid) and the $g(a)$ function arising in
    the NLO free-streaming expansion (dashed) for $w_0$CDM models with
    various $w_0$.  Note that, since $\Om(a)$, $G(a)$, and $dD(a)/da$ of
    \EQEQEQ{e:Oma_Oda_w0cdm}{e:G_w0cdm}{e:dDda_w0cdm} can be written as
    functions of $a^{-w_0}/r_0^{1/3}$ and $w_0$ alone, our choice of time
    variables on the $x$ axis scales out all other cosmological parameter
    dependence.  For $\Omo$ of $0.1$, $0.2$, and $0.33$, the present day
    ($a=1$) corresponds to $a^{-w_0}/r_0^{1/3}$ of $\approx 2.08$, $1.59$, and
    $1.27$, respectively.
    \label{f:g_nlo_vs_1-G}
  }
\end{figure}

Comparing the NLO fit of \EQ{e:fit_nlo} to the \SEEdS{} fit of \EQ{e:fit_eul}, we see that the function $1-G(a)$ describing deviations from EdS in \SEEdS{} is replaced by $g(a)$ in the NLO fit.  Figure~\ref{f:g_nlo_vs_1-G} compares these two for a range of $a$ and $w_0$.  Remarkably, at early times $a^{-w_0}/r_0^{1/3} \ll 1$, the two agree to $<7\%$ for $w_0=-1$.  Thus, in this case, \SEEdS{} is a close approximation to the NLO fit.  The  fits disagree qualitatively in two ways.  Firstly, \SEEdS{} depends much more strongly upon $w_0$.  At $a^{-w_0}/r_0 = 0.1$, $1-G(a)$ varies by a factor of two between $w_0$ of $-1$ and $-2$, but $g(a)$ varies by only $4\%$.  Secondly, at late times, $1-G(a)$ is bounded from above by $1$, while the smallness of $\Om(a)$ allows $g(a)$ to be considerably larger.  

\begin{figure}[tb]
  \hskip-2mm\includegraphics[width=77mm]{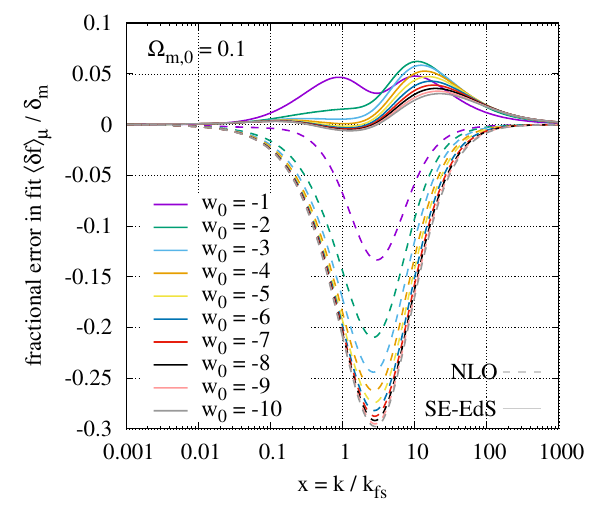}
  ~\includegraphics[width=77mm]{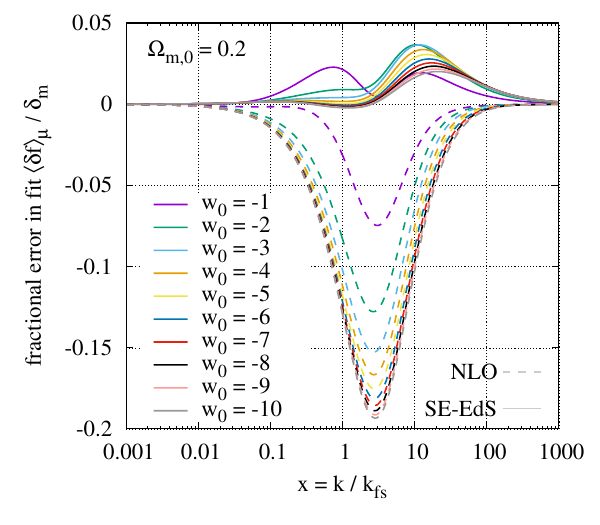}
  \caption{
    Comparison of the \SEEdS{} fit of \EQ{e:fit_eul} (solid)
    and the NLO fit of \EQ{e:fit_nlo} (dashed) at $a=1$ and $p/m_\nu=0.01$
    for $w_0$CDM models with
    (Left)~$\Omo=0.1$ and
    (Right)~$\Omo=0.2$. All other parameters have been set
    to their fiducial values of \EQ{e:fid_cosmo}.  While the two fits have
    comparable errors for $w_0=-1$, the NLO fit is badly behaved for large
    negative $w_0$ and small $\Omo$.
    \label{f:vary_w0_SEEdS_vs_NLO}
  }
\end{figure}

This comparison of $1-G$ and $g$ suggests testing \SEEdS{} against the NLO fit at low $\Omo$ and very negative $w_0$.  Figure~\ref{f:vary_w0_SEEdS_vs_NLO} performs this comparison for $\Omo$ of $0.1$ and $0.2$, reducing $w_0$ as low as $-10$.  The \SEEdS{} errors saturate around $w_0 = -3$, then drop slowly as $w_0$ is reduced further, as noticed earlier in Fig.~\ref{f:eul_vary_Onu0_w0}~(Right).  However, the NLO errors continue to grow as $w_0$ is reduced to $w_0=-10$, reaching $30\%$ for $\Omo=0.1$ and nearly $20\%$ for $\Omo=0.2$.  Whereas $1-G$ falls $\propto |w_0|^{-1}$ at large negative $w_0$, as shown in \EQ{e:G_large_w0}, respecting all three EdS limits of Sec.~\ref{sub:fit:3limits}, the NLO fit does not approach EdS at small $|w_0|^{-1}$, leading to the large errors seen in Fig.~\ref{f:vary_w0_SEEdS_vs_NLO}.  Thus \SEEdS{} is effectively a generalization of the NLO fit to a larger range of dark energy models.

\subsection{Lagrangian-momentum free-streaming expansion}
\label{sub:nlo:lag}

The Lagrangian-momentum case proceeds similarly.  From  \EQ{e:delta_a_k_u_int_y}, we define $\ILc_0(k,y)$ as
\begin{equation}
  \ILc_0(k,y) := \int_y^{\yin} dy' a(y')^2 \Phi(y',k) \sin(y'-y)
  \quad\Rightarrow\quad
  \delta(a,k,u) = -\frac{1}{u^2} \ILc_0(k,y).
\end{equation}
Integrating by parts with $\ILc_0 = \int \ULc_0 d\VLc_0$, $\ULc_0(y') = a(y')^2 \Phi(y')$, and $\VLc_0(y') = -\cos(y'-y)$ gives $\ILc_0 = \left.[\ULc_0(y') \VLc_0(y')]\right|_y^{\yin} - \int \VLc_0 d\ULc_0$.  This is precisely the result of \EQ{e:delta_a_k_u_int_by_parts}, whose boundary term in the $\yin \rightarrow \infty$ limit is the leading order behavior $\delta/\dm \rightarrow 1/x^2$.

Proceeding, we define $\ILc_1 = - \int \VLc_0 d\ULc_0 = \int \ULc_1 d\VLc_1$, with $\ULc_1(y') = \partial(a^2\Phi)/\partial y'$ and $\VLc_1(y') = \sin(y'-y)$.  However, in this case the boundary term $\left.[\ULc_1(y') \VLc_1(y')]\right|_y^{\yin}$ vanishes, since $\sin(y'-y)$ is zero at $y'=y$ and $a(y')\rightarrow 0$ as $\yin \rightarrow \infty$.  Defining $\ILc_2 = \int \VLc_1 d\ULc_1 = \int \ULc_2 d\VLc_2$ with $\ULc_2 = \partial^2(a^2\Phi)/\partial(y')^2$ and $\VLc_2 = \cos(y'-y)$, we have $\ILc_2 = \left.[ \ULc_2 \VLc_2 ]\right|_y^{\yin} - \int \VLc_2 d\ULc_2$, and so forth.  The greater simplicity of the window function $\sin(y'-y)$ compared with $W(y'-y)$ in the Eulerian case makes repeated integration by parts simple.  We have $\ULc_n = \partial^n (a^2 \Phi) / \partial(y')^n$, $\VLc_{2n} = (-1)^{n+1} \cos(y'-y)$, and $\VLc_{2n+1} = (-1)^n \sin(y'-y)$.  Boundary terms vanish for odd $n$, and for even $n$ are $\left.[\ULc_{2n} \VLc_{2n}]\right|_y^{\yin} \rightarrow (-1)^n \partial^{2n} (a(y)^2 \Phi(y)) / \partial y^{2n}$.  The full expansion is
\begin{equation}
  \ILc_0(k,y) = \sum_{n=0}^\infty (-1)^n \left(\frac{\partial}{\partial y}\right)^{2n}\!\!\left[a(y)^2 \Phi(k,y)\right]
  \label{e:lag_fs_full_expansion}
\end{equation}

In the linear EdS case, \EQ{e:da2Phi_dy_EdS_lin} shows that each new derivative of $a^2\Phi$ with respect to $y$ introduces an additional power of $x^{-1}$.  Immediately, we notice that the Lagrangian-momentum free-streaming expansion has only even powers of $x^{-1}$.  This should not be surprising, as the high-$x$ asymptotic expansion of the exact EdS solution of \EQ{e:dhdm_dm_eds_lag} also only has even terms.  However, it is qualitatively different from the Eulerian expansion, and is incompatible with the \SEEdS{} fit of \EQ{e:fit_lag}, whose NLO high-$x$ term is $\propto x^{-3}$.

Nevertheless, let us investigate the possibility of a new rational function fit based upon the NLO expansion of \EQ{e:lag_fs_full_expansion}, that is, the $n=0$ and $n=1$ terms.  In linear theory, where the time-dependence of $\Phi$ is simply proportional to $G$, the NLO expansion for $\delta/\dm$ is
\begin{equation}
  \frac{\delta(a,k,u)}{\dm(a,k)} \approx \frac{1}{x^2} - \frac{\cL}{x^4}
  \quad\mathrm{where}\quad
  \cL = \frac{2}{3 \Omo \Hco^2} \frac{1}{a^3 G} \frac{\partial^2}{\partial s^2}[a(s)^2 \Phi(s,k)].
\end{equation}
In EdS, $\cL$ is simply $10/3$.  Near our fiducial $\Lambda$CDM model of \EQ{e:fid_cosmo}, the approximation $\Om(a) \approx f(a)^{5/9}$, with $f(a) = d \log(D) / d \log(a)$, is accurate to a few percent, and we find $\cL \approx 7.5$ at $a=1$ for the fiducial case.  Requiring that a fitting formula approach the ratio of the $\Lambda$CDM to EdS free-streaming limits at high $x$, and choosing the minimal rational function interpolating between this and the clustering limit, we find
\begin{equation}
  \Fc(a,x) = \frac{b_0 + x^3}{b_0 + b_1 x + x^3}
  \quad\mathrm{where}\quad
  b_1 = \cL - \frac{10}{3}.
  \label{e:lag_nlo_fit_attempt}
\end{equation}
A \SEEdS{}-like fit would set $b_0$ to $3$, but we keep it as a free parameter here.  

We find that \EQ{e:lag_nlo_fit_attempt} for $b_0 = 3$ severely underpredicts the actual $\delta/\dm$, by about $45\%$.  Scanning across the range $1 \leq b_0 \leq 10$, we find maximum errors from $65\%$ to $25\%$.  Although much larger $b_0$, $60 \lesssim b_0 \lesssim 100$, can reduce errors to $5\%$, at no $b_0$ does it approach the percent-level accuracy of \SEEdS{} for the fiducial model.  It is also not clear how such a large $b_0$ could be predicted from first principles.  Phenomenologically, a large $b_0$ dials down the impact of the NLO correction term $b_1 x$ in the interpolation region $0.1 \lesssim x \lesssim 10$, where errors tend to be largest.  Thus our quest for a more rigorous alternative to $\FL(a,x)$ through higher-order corrections faces defeat.  Even the best-motivated correction, the high-$x$ NLO correction, requires that a parameter whose value we cannot predict be tuned to a value $\gg 1$ precisely to suppress the effect of that NLO correction.

\begin{figure}[tb]
  \begin{center}
    \includegraphics[width=140mm]{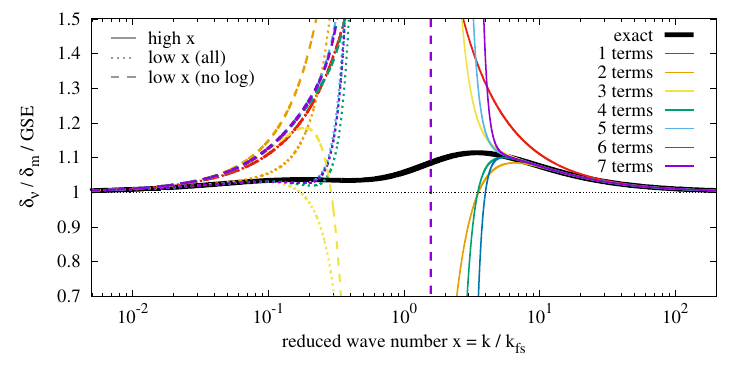}%
  \end{center}
  \caption{
    Exact EdS solution of \EQ{e:dhdm_dm_eds_lag} (thick black line) as well as
    high-$x$ asymptotic expansions of \EQ{e:dnu_dm_eds_high_x} (thin solid
    lines), low-$x$ expansions of \EQ{e:dnu_dm_eds_low_x}
    (dotted lines), and low-$x$ expansions excluding the non-perturbative
    terms $\propto \log(x\sqrt{6})$ (dashed lines). Leading-order expansions
    (1 term), NLO (2 terms), and several higher orders are shown.
    \label{f:EdS_asymp}
  }
\end{figure}

This suggests that higher-order corrections, even the NLO term $\propto x^{-4}$, are not the optimal approach to fitting $\RL$ in the interpolation region.  We  investigate this question in greater detail for the EdS solution of \EQ{e:dhdm_dm_eds_lag}, which we know exactly and which may be expanded to arbitrary numbers of terms in both the low-$x$ and high-$x$ limits:
\begin{eqnarray}
  \left.\frac{\delta}{\dm}\right|_{{\rm EdS,}x\rightarrow0}
  &=&
  1 + 6[\log(x\sqrt{6}) \!+\! \gammaEM]x^2 - 3\sqrt{6}\pi x^3
  + 18[\log(x\sqrt{6}) \!+\! \tfrac{3}{2} \!+\! \gammaEM]x^4 + 3\sqrt{6}\pi x^5 
  \ldots\nonumber\\
  &~&
  + \tfrac{3}{4}[12\log(x\sqrt{6}) \!+\! 12\gammaEM \!-\! 25]x^6
  - \tfrac{9\sqrt{6}\pi}{10}x^7
  + {\mathcal O}(x^8 \log(x))
  \label{e:dnu_dm_eds_low_x}
  \\
  \left.\frac{\delta}{\dm}\right|_{{\rm EdS,}x\rightarrow\infty}
  &=&
  \frac{1}{x^2} - \frac{10}{3x^4} + \frac{70}{3x^6} - \frac{280}{x^8}
  + \frac{15400}{3x^{10}} - \frac{400400}{3x^{12}} + \frac{14014000}{3x^{14}}
  + {\mathcal O}(x^{-16})
  \label{e:dnu_dm_eds_high_x}
\end{eqnarray}
Figure~\ref{f:EdS_asymp} compares the exact solution to these expansions including varying numbers of terms.  For the purpose of the plot, in the low-$x$ expansion of \EQ{e:dnu_dm_eds_low_x}, a given power of $x$ multiplying any number of quantities in square brackets is counted as a single term.  That is, we count \EQ{e:dnu_dm_eds_low_x} as containing seven terms, and the expansion to two terms is $1 + 6[\log(x\sqrt{6}) + \gammaEM]x^2$.  The figure also shows the low-$x$ expansion omitting all of the $\log(x\sqrt{6})$ quantities, which will not arise in a power series expansion about $x=0$.  

Figure~\ref{f:EdS_asymp} illustrates two of the difficulties with higher-order expansions.  Firstly, although NLO ($2$ terms) and higher-order expansions do improve agreement outside of the interpolation region, this comes at the cost of much quicker divergence from the exact solution within the interpolation region.  This imposes greater difficulties upon a rational function such as \EQ{e:lag_nlo_fit_attempt} interpolating between the clustering and free-streaming limits, causing it to raise coefficients such as $b_0$ to cancel the bad behavior of higher-order terms within the interpolation region.  Secondly, at low $x$, both exact EdS solutions, \EQEQ{e:psieds}{e:dhdm_dm_eds_lag}, have quantities $\propto \log(x \sqrt{6})$ at NLO and higher orders.  Omitting these terms, which will not arise in a Taylor expansion about $x=0$, makes the NLO fit worse than the leading order even outside of the interpolation region, $x \lesssim 0.1$.

The lesson of this Section, then, is that while higher-order corrections in the free-streaming limit provide valuable guidance towards fitting functions for $\dnu/\dm$, they do not as yet supply a broadly-applicable, theoretically-rigorous fit within the interpolation region, where errors are largest.  This is expecially true in the Lagrangian momentum case, for which the NLO correction contains two additional powers of $x^{-1}$ relative to the leading-order free-streaming limit.  Our best fitting functions, the \SEEdS{} fits of \EQEQ{e:fit_eul}{e:fit_lag}, are phenomenological fits guided by the three EdS limits of Sec.~\ref{sub:fit:3limits}.

\section{Conclusions}
\label{sec:con}

The chief results of this Article are the two \SEEdS{} fitting functions, \EQ{e:fit_eul} for the Eulerian $\dfmu(a,k,p)/\dm(a,k)$ and \EQ{e:fit_lag} for the Lagrangian-momentum $\delta(a,k,u)/\dm(a,k)$, relative to their EdS equivalents, and their integrated forms, \EQEQ{e:dnu_tot_eul}{e:dnu_tot_lag}.  By design, these fits approach unity when: $k/\kfs \rightarrow 0$; $k/\kfs \rightarrow \infty$; $\Omo \rightarrow 1$; $a \ll 1$ when radiation is neglected; and $w_0 \rightarrow -\infty$, which brings the scale factor of matter-dark-energy-equality arbitrarily close to unity.  The time-dependence of each fit occurs entirely through the large-scale reduced growth factor $G(a) = D(a)/a$.  We have confirmed their accuracy to $<3\%$ for a wide variety of cosmological parameters: $0.2 \leq \Omo \leq 1$; $\Ono \leq 0.018$, equivalent to $\Mnu \leq 730$~meV for our fidicual model; a constant $w_0 \leq -0.6$; and a curvature $|\OKo| \leq 0.1$.  The \SEEdS{} fits are accurate for a broad range of non-clustering CPL dark energies, aside from the early dark energies whose density fractions $\gtrsim 0.5\%$ at $a=10^{-3}$ lead to large departures from EdS over an extended period.  The fits are even accurate for modified growth rates $d\log(D)/d\log(a) = \Om(a)^\gamma$ with $0.4 \leq \gamma \leq 0.9$.  

One  application of our fits is as an inexpensive, easily-implemented linear response approximation to neutrino clustering in N-body simulations.  Section~\ref{sub:fit:paint} outlines a \SEEdS{}-based linear response in which neutrinos are painted into a simulation. Testing this application using an emulator for the matter clustering, we find it to be considerably more accurate than competing fitting functions in the linear regime, $k \leq 0.1~h/$Mpc.  In the non-linear regime, it slightly overestimates linear response neutrinos precisely where linear response more significantly underestimates non-linear neutrinos, making the error in the \SEEdS{} fits subdominant.  Moreover, \SEEdS{} fits provide information on the dependence of neutrino clustering upon either Eulerian or Lagrangian momenta.   This momentum resolution provides insights into the distribution-function-dependence of neutrino constraints, allows for the inclusion of other HDM in addition to neutrinos, and facilitates laboratory constrains on the cosmic neutrino background, which depend upon the local momentum as well as the density of neutrinos.

Finally, we explored the close correspondence between our Eulerian-momentum \SEEdS{} fit and higher-order free-streaming corrections.  For $\nu\Lambda$CDM at early times, the two corrections agree to several percent.  At late times, and especially for very negative dark energy equations of state, \SEEdS{} is more accurate.  Thus \SEEdS{} is a generalization of the next-to-leading-order corrections which evades their divergent behavior for $w_0 \rightarrow -\infty$ and low $\Om(a)$.  We further discussed the failure of higher-order expansions for $k \lesssim \kfs$ due to non-perturbative logarithmic terms, further motivating the \SEEdS{} fits.   

In conclusion, the \SEEdS{} fits of \EQEQ{e:fit_eul}{e:fit_lag} are fast, accurate, and inexpensive approximations to the linear clustering of neutrinos and other HDM.  They are thus ideal for applications in which linear neutrinos are sufficient, such as simulations quantifying the effects of neutrinos upon galaxy clustering.

\subsection*{Acknowledgments}

The authors are grateful to J.~Kwan, I.~G.~McCarthy, M.~Mosbech, and L.~Tan for insightful discussions.  AU is supported by the National Natural Science Foundation of China (NSFC) under grant No.~12473005 as well as the Yunnan Key Laboratory of Survey Science under Project No.~202449CE340002.  Y$^3$W is supported by the Australian Research Council’s Discovery Project (DP240103130) scheme.

\bibliographystyle{bibi}
\bibliography{seeds}

\providecommand{\href}[2]{#2}\begingroup\raggedright\begin{thebibliography}{10}

\bibitem{Planck:2018vyg}
{\scshape Planck} Collaboration, N.~Aghanim et~al., \emph{{Planck 2018 results.
  VI. Cosmological parameters}},
  \href{https://doi.org/10.1051/0004-6361/201833910}{\emph{Astron. Astrophys.}
  {\bfseries 641} (2020) A6}
  [\href{https://arxiv.org/abs/1807.06209}{{\ttfamily 1807.06209}}]. [Erratum:
  Astron.Astrophys. 652, C4 (2021)].

\bibitem{DES:2024jxu}
{\scshape DES} Collaboration, T.~M.~C. Abbott et~al., \emph{{The Dark Energy
  Survey: Cosmology Results with {\ensuremath{\sim}}1500 New High-redshift Type
  Ia Supernovae Using the Full 5 yr Data Set}},
  \href{https://doi.org/10.3847/2041-8213/ad6f9f}{\emph{Astrophys. J. Lett.}
  {\bfseries 973} (2024) L14}
  [\href{https://arxiv.org/abs/2401.02929}{{\ttfamily 2401.02929}}].

\bibitem{DESI:2024hhd}
{\scshape DESI} Collaboration, A.~G. Adame et~al., \emph{{DESI 2024 VII:
  cosmological constraints from the full-shape modeli ng of clustering
  measurements}},
  \href{https://doi.org/10.1088/1475-7516/2025/07/028}{\emph{JCAP} {\bfseries
  07} (2025) 028} [\href{https://arxiv.org/abs/2411.12022}{{\ttfamily
  2411.12022}}].

\bibitem{DESI:2025ejh}
{\scshape DESI} Collaboration, W.~Elbers et~al., \emph{{Constraints on Neutrino
  Physics from DESI DR2 BAO and DR1 Full Shape}},
  {\emph{e-Print:arXiv:2503.14744} (2025) }
  [\href{https://arxiv.org/abs/2503.14744}{{\ttfamily 2503.14744}}].

\bibitem{SPT-3G:2025bzu}
{\scshape SPT-3G} Collaboration, E.~Camphuis et~al., \emph{{SPT-3G D1: CMB
  temperature and polarization power spectra and cosmology from 2019 and 2020
  observations of the SPT-3G main field}},
  \href{https://doi.org/10.1103/7wt3-9v2y}{\emph{Phys. Rev. D} {\bfseries 113}
  (2026) 083504} [\href{https://arxiv.org/abs/2506.20707}{{\ttfamily
  2506.20707}}].

\bibitem{DayaBay:2022orm}
{\scshape Daya Bay} Collaboration, F.~P. An et~al., \emph{{Precision
  Measurement of Reactor Antineutrino Oscillation at Kilometer-Scale Baselines
  by Daya Bay}},
  \href{https://doi.org/10.1103/PhysRevLett.130.161802}{\emph{Phys. Rev. Lett.}
  {\bfseries 130} (2023) 161802}
  [\href{https://arxiv.org/abs/2211.14988}{{\ttfamily 2211.14988}}].

\bibitem{Esteban:2024eli}
I.~Esteban, M.~C. Gonzalez-Garcia, M.~Maltoni, I.~Martinez-Soler, J.~P.
  Pinheiro and T.~Schwetz, \emph{{NuFit-6.0: updated global analysis of
  three-flavor neutrino oscillations}},
  \href{https://doi.org/10.1007/JHEP12(2024)216}{\emph{JHEP} {\bfseries 12}
  (2024) 216} [\href{https://arxiv.org/abs/2410.05380}{{\ttfamily
  2410.05380}}].

\bibitem{JUNO:2025gmd}
{\scshape JUNO} Collaboration, A.~Abusleme et~al., \emph{{Measurement of
  reactor neutrino oscillation with the first JUNO data}},
  \href{https://doi.org/10.1038/s41586-026-10538-z}{\emph{Nature} {\bfseries
  654} (2026) 343} [\href{https://arxiv.org/abs/2511.14593}{{\ttfamily
  2511.14593}}].

\bibitem{Ong:2026tta}
D.~D.~Y. Ong, D.~Yallup and W.~Handley, \emph{{The Bayesian view of DESI DR2
  with unimpeded: Evidence and tension in a combined analysis with CMB and
  supernovae across cosmological models}},
  \href{https://arxiv.org/abs/2603.05472}{{\ttfamily 2603.05472}}.

\bibitem{Karim:2026jcx}
T.~Karim, J.~S. Speagle and R.~Hlo{\v{z}}ek, \emph{{Model Predictive Scoring
  Shows Specific BAO Observations (not SNIa) Drives $w_0w_a$ Tension}},
  \href{https://arxiv.org/abs/2609.04334}{{\ttfamily 2609.04334}}.

\bibitem{Cai:2026swf}
R.-G. Cai and S.-J. Wang, \emph{{The Hubble Tension: A Decade Review}},
  \href{https://doi.org/10.1088/1674-4527/ae842f}{\emph{Res. Astron.
  Astrophys.} {\bfseries 26} (2026) 084011}
  [\href{https://arxiv.org/abs/2606.20434}{{\ttfamily 2606.20434}}].

\bibitem{Efstathiou:2026fax}
G.~Efstathiou, \emph{{Late Time Dynamical Dark Energy and the CMB-Distance
  Ladder Tension}},  \href{https://arxiv.org/abs/2607.29562}{{\ttfamily
  2607.29562}}.

\bibitem{Li:2026asg}
T.-N. Li, G.-H. Du, H.~Wang, Y.-H. Li, J.-F. Zhang and X.~Zhang, \emph{{Dark
  Energy in the DESI Era: A Brief Review of Evidence,
  Beyond-{\ensuremath{\Lambda}}CDM Interpretations, and Tensions}},
  \href{https://doi.org/10.1088/1674-4527/ae8429}{\emph{Res. Astron.
  Astrophys.} {\bfseries 26} (2026) 084002}
  [\href{https://arxiv.org/abs/2606.21826}{{\ttfamily 2606.21826}}].

\bibitem{Schoneberg:2026vaf}
N.~Sch{\"o}neberg, V.~Poulin, A.~G. Ferrari, F.~Finelli, J.~Lesgourgues,
  L.~Morelli, M.~R. Mosbech, R.~K. Sharma and T.~Simon, \emph{{The $H_0$ world
  cup. II. A comprehensive competition between proposed Hubble tension
  solutions}},  \href{https://arxiv.org/abs/2607.13283}{{\ttfamily
  2607.13283}}.

\bibitem{Freedman:2024eph}
W.~L. Freedman, B.~F. Madore, T.~J. Hoyt, I.~S. Jang, A.~J. Lee and K.~A.
  Owens, \emph{{Status Report on the Chicago-Carnegie Hubble Program (CCHP):
  Measurement of the Hubble Constant Using the Hubble and James Webb Space
  Telescopes}},
  \href{https://doi.org/10.3847/1538-4357/adce78}{\emph{Astrophys. J.}
  {\bfseries 985} (2025) 203}
  [\href{https://arxiv.org/abs/2408.06153}{{\ttfamily 2408.06153}}]. [Erratum:
  Astrophys.J. 993, 252 (2025)].

\bibitem{Leauthaud:2016jdb}
A.~Leauthaud et~al., \emph{{Lensing is Low: Cosmology, Galaxy Formation, or New
  Physics?}}, \href{https://doi.org/10.1093/mnras/stx258}{\emph{Mon. Not. Roy.
  Astron. Soc.} {\bfseries 467} (2017) 3024}
  [\href{https://arxiv.org/abs/1611.08606}{{\ttfamily 1611.08606}}].

\bibitem{Zu:2025lrk}
L.~Zu, W.~Giar{\`e}, C.~Zhang, E.~Di~Valentino, Y.-L.~S. Tsai and
  S.~Trojanowski, \emph{{A solution to the S$_{8}$ tension through
  neutrino{\textendash}dark matter interactions}},
  \href{https://doi.org/10.1038/s41550-025-02733-1}{\emph{Nature Astron.}
  {\bfseries 10} (2026) 457}
  [\href{https://arxiv.org/abs/2501.13785}{{\ttfamily 2501.13785}}].

\bibitem{Pantos:2026koc}
I.~Pantos and L.~Perivolaropoulos, \emph{{Status of the $S_8$ Tension: A 2026
  Review of Probe Discrepancies}},
  \href{https://doi.org/10.1016/j.dark.2026.102286}{\emph{Phys. Dark Univ.}
  {\bfseries 52} (2026) 102286}
  [\href{https://arxiv.org/abs/2602.12238}{{\ttfamily 2602.12238}}].

\bibitem{DiValentino:2021izs}
E.~Di~Valentino, O.~Mena, S.~Pan, L.~Visinelli, W.~Yang, A.~Melchiorri, D.~F.
  Mota, A.~G. Riess and J.~Silk, \emph{{In the realm of the Hubble
  tension{\textemdash}a review of solutions}},
  \href{https://doi.org/10.1088/1361-6382/ac086d}{\emph{Class. Quant. Grav.}
  {\bfseries 38} (2021) 153001}
  [\href{https://arxiv.org/abs/2103.01183}{{\ttfamily 2103.01183}}].

\bibitem{Kamionkowski:2022pkx}
M.~Kamionkowski and A.~G. Riess, \emph{{The Hubble Tension and Early Dark
  Energy}},
  \href{https://doi.org/10.1146/annurev-nucl-111422-024107}{\emph{Ann. Rev.
  Nucl. Part. Sci.} {\bfseries 73} (2023) 153}
  [\href{https://arxiv.org/abs/2211.04492}{{\ttfamily 2211.04492}}].

\bibitem{Jhaveri:2025neg}
T.~Jhaveri, T.~Karwal and W.~Hu, \emph{{Turning a negative neutrino mass into a
  positive optical depth}},
  \href{https://doi.org/10.1103/6vd2-rbfn}{\emph{Phys. Rev. D} {\bfseries 112}
  (2025) 043541} [\href{https://arxiv.org/abs/2504.21813}{{\ttfamily
  2504.21813}}].

\bibitem{Sailer:2025lxj}
N.~Sailer, G.~S. Farren, S.~Ferraro and M.~White, \emph{{Dispu$\tau$able: the
  high cost of a low optical depth}}, {\emph{e-Print:arXiv:2504.16932} (2025) }
  [\href{https://arxiv.org/abs/2504.16932}{{\ttfamily 2504.16932}}].

\bibitem{Brandbyge:2009ce}
J.~Brandbyge and S.~Hannestad, \emph{{Resolving Cosmic Neutrino Structure: A
  Hybrid Neutrino N-body Scheme}},
  \href{https://doi.org/10.1088/1475-7516/2010/01/021}{\emph{JCAP} {\bfseries
  01} (2010) 021} [\href{https://arxiv.org/abs/0908.1969}{{\ttfamily
  0908.1969}}].

\bibitem{Banerjee:2018bxy}
A.~Banerjee, D.~Powell, T.~Abel and F.~Villaescusa-Navarro, \emph{{Reducing
  Noise in Cosmological N-body Simulations with Neutrinos}},
  \href{https://doi.org/10.1088/1475-7516/2018/09/028}{\emph{JCAP} {\bfseries
  09} (2018) 028} [\href{https://arxiv.org/abs/1801.03906}{{\ttfamily
  1801.03906}}].

\bibitem{Elbers:2020lbn}
W.~Elbers, C.~S. Frenk, A.~Jenkins, B.~Li and S.~Pascoli, \emph{{An optimal
  non-linear method for simulating relic neutrinos}},
  \href{https://doi.org/10.1093/mnras/stab2260}{\emph{Mon. Not. Roy. Astron.
  Soc.} {\bfseries 507} (2021) 2614}
  [\href{https://arxiv.org/abs/2010.07321}{{\ttfamily 2010.07321}}].

\bibitem{Chen:2022cgw}
J.~Z. Chen, A.~Upadhye and Y.~Y.~Y. Wong, \emph{{Flows for the masses: A
  multi-fluid non-linear perturbation theory for massive neutrinos}},
  \href{https://doi.org/10.1088/1475-7516/2023/05/046}{\emph{JCAP} {\bfseries
  05} (2023) 046} [\href{https://arxiv.org/abs/2210.16020}{{\ttfamily
  2210.16020}}].

\bibitem{Upadhye:2024ypg}
A.~Upadhye, M.~R. Mosbech, G.~Pierobon and Y.~Y.~Y. Wong, \emph{{Everything hot
  everywhere all at once: neutrinos and hot dark matter as a single effective
  species}}, \href{https://doi.org/10.1088/1475-7516/2025/01/077}{\emph{JCAP}
  {\bfseries 01} (2025) 077}
  [\href{https://arxiv.org/abs/2410.05815}{{\ttfamily 2410.05815}}].

\bibitem{Garny:2022fsh}
M.~Garny and P.~Taule, \emph{{Two-loop power spectrum with full time- and
  scale-dependence and EFT corrections: impact of massive neutrinos and going
  beyond EdS}},
  \href{https://doi.org/10.1088/1475-7516/2022/09/054}{\emph{JCAP} {\bfseries
  09} (2022) 054} [\href{https://arxiv.org/abs/2205.11533}{{\ttfamily
  2205.11533}}].

\bibitem{Brandbyge:2008js}
J.~Brandbyge and S.~Hannestad, \emph{{Grid Based Linear Neutrino Perturbations
  in Cosmological N-body Simulations}},
  \href{https://doi.org/10.1088/1475-7516/2009/05/002}{\emph{JCAP} {\bfseries
  05} (2009) 002} [\href{https://arxiv.org/abs/0812.3149}{{\ttfamily
  0812.3149}}].

\bibitem{Ali-Haimoud:2012fzp}
Y.~Ali-Haimoud and S.~Bird, \emph{{An efficient implementation of massive
  neutrinos in non-linear structure formation simulations}},
  \href{https://doi.org/10.1093/mnras/sts286}{\emph{Mon. Not. Roy. Astron.
  Soc.} {\bfseries 428} (2012) 3375}
  [\href{https://arxiv.org/abs/1209.0461}{{\ttfamily 1209.0461}}].

\bibitem{Ji:2022iji}
L.~Ji, M.~Kamionkowski and J.~L. Bernal, \emph{{Cosmological perturbations:
  Noncold relics without the Boltzmann hierarchy}},
  \href{https://doi.org/10.1103/PhysRevD.106.103531}{\emph{Phys. Rev. D}
  {\bfseries 106} (2022) 103531}
  [\href{https://arxiv.org/abs/2201.11129}{{\ttfamily 2201.11129}}].

\bibitem{Lee:2025zym}
N.~Lee, J.~L. Bernal, S.~G{\"u}nther, L.~Ji and M.~Kamionkowski, \emph{{Rapid
  and accurate numerical evolution of linear cosmological perturbations with
  noncold relics}}, \href{https://doi.org/10.1103/fs6c-zbzf}{\emph{Phys. Rev.
  D} {\bfseries 112} (2025) 043529}
  [\href{https://arxiv.org/abs/2506.01956}{{\ttfamily 2506.01956}}].

\bibitem{Lee:2025vgv}
N.~Lee, Y.~Ali-Ha{\"\i}moud and M.~Kamionkowski, \emph{{Efficient analytic
  approximation for small-scale noncold relic perturbations}},
  \href{https://doi.org/10.1103/v219-d6tz}{\emph{Phys. Rev. D} {\bfseries 113}
  (2026) 043525} [\href{https://arxiv.org/abs/2510.20821}{{\ttfamily
  2510.20821}}].

\bibitem{Ringwald:2004np}
A.~Ringwald and Y.~Y.~Y. Wong, \emph{{Gravitational clustering of relic
  neutrinos and implications for their detection}},
  \href{https://doi.org/10.1088/1475-7516/2004/12/005}{\emph{JCAP} {\bfseries
  12} (2004) 005} [\href{https://arxiv.org/abs/hep-ph/0408241}{{\ttfamily
  hep-ph/0408241}}].

\bibitem{Wong:2008ws}
Y.~Y.~Y. Wong, \emph{{Higher order corrections to the large scale matter power
  spectrum in the presence of massive neutrinos}},
  \href{https://doi.org/10.1088/1475-7516/2008/10/035}{\emph{JCAP} {\bfseries
  10} (2008) 035} [\href{https://arxiv.org/abs/0809.0693}{{\ttfamily
  0809.0693}}].

\bibitem{Chen:2020kxi}
J.~Z. Chen, A.~Upadhye and Y.~Y.~Y. Wong, \emph{{One line to run them all:
  SuperEasy massive neutrino linear response in $N$-body simulations}},
  \href{https://doi.org/10.1088/1475-7516/2021/04/078}{\emph{JCAP} {\bfseries
  04} (2021) 078} [\href{https://arxiv.org/abs/2011.12504}{{\ttfamily
  2011.12504}}].

\bibitem{Pierobon:2024kpw}
G.~Pierobon, M.~R. Mosbech, A.~Upadhye and Y.~Y.~Y. Wong, \emph{{One trick to
  treat them all: SuperEasy linear response for any hot dark matter in N-body
  simulations}},
  \href{https://doi.org/10.1088/1475-7516/2024/12/032}{\emph{JCAP} {\bfseries
  12} (2024) 032} [\href{https://arxiv.org/abs/2410.05816}{{\ttfamily
  2410.05816}}].

\bibitem{Dupuy:2013jaa}
H.~Dupuy and F.~Bernardeau, \emph{{Describing massive neutrinos in cosmology as
  a collection of independent flows}},
  \href{https://doi.org/10.1088/1475-7516/2014/01/030}{\emph{JCAP} {\bfseries
  01} (2014) 030} [\href{https://arxiv.org/abs/1311.5487}{{\ttfamily
  1311.5487}}].

\bibitem{Dupuy:2014vea}
H.~Dupuy and F.~Bernardeau, \emph{{Cosmological Perturbation Theory for streams
  of relativistic particles}},
  \href{https://doi.org/10.1088/1475-7516/2015/03/030}{\emph{JCAP} {\bfseries
  03} (2015) 030} [\href{https://arxiv.org/abs/1411.0428}{{\ttfamily
  1411.0428}}].

\bibitem{Lee:2009if}
S.~Lee and K.-W. Ng, \emph{{Growth index in an accelerating universe}},
  {\emph{Chin. J. Phys.} {\bfseries 50} (2012) 367}
  [\href{https://arxiv.org/abs/0905.1522}{{\ttfamily 0905.1522}}].

\bibitem{Lee:2009gb}
S.~Lee and K.-W. Ng, \emph{{Growth index with the exact analytic solution of
  sub-horizon scale linear perturbation for dark energy models with constant
  equation of state}},
  \href{https://doi.org/10.1016/j.physletb.2010.03.082}{\emph{Phys. Lett. B}
  {\bfseries 688} (2010) 1} [\href{https://arxiv.org/abs/0906.1643}{{\ttfamily
  0906.1643}}].

\bibitem{Lee:2009ft}
S.~Lee and K.-W. Ng, \emph{{Properties of the exact analytic solution of the
  growth factor and its applications}},
  \href{https://doi.org/10.1103/PhysRevD.82.043004}{\emph{Phys. Rev. D}
  {\bfseries 82} (2010) 043004}
  [\href{https://arxiv.org/abs/0907.2108}{{\ttfamily 0907.2108}}].

\bibitem{Upadhye:2025glk}
A.~Upadhye and Y.~Li, \emph{{Slow neutrinos: non-linearity and momentum-space
  emulation}}, \href{https://doi.org/10.1093/mnras/stag382}{\emph{Mon. Not.
  Roy. Astron. Soc.} {\bfseries 547} (2026) stag382}
  [\href{https://arxiv.org/abs/2511.09945}{{\ttfamily 2511.09945}}].

\bibitem{Linder:2003dr}
E.~V. Linder and A.~Jenkins, \emph{{Cosmic structure and dark energy}},
  \href{https://doi.org/10.1046/j.1365-2966.2003.07112.x}{\emph{Mon. Not. Roy.
  Astron. Soc.} {\bfseries 346} (2003) 573}
  [\href{https://arxiv.org/abs/astro-ph/0305286}{{\ttfamily
  astro-ph/0305286}}].

\bibitem{Ma:1995ey}
C.-P. Ma and E.~Bertschinger, \emph{{Cosmological perturbation theory in the
  synchronous and conformal Newtonian gauges}},
  \href{https://doi.org/10.1086/176550}{\emph{Astrophys. J.} {\bfseries 455}
  (1995) 7} [\href{https://arxiv.org/abs/astro-ph/9506072}{{\ttfamily
  astro-ph/9506072}}].

\bibitem{Chen:2020bdf}
J.~Z. Chen, A.~Upadhye and Y.~Y.~Y. Wong, \emph{{The cosmic neutrino background
  as a collection of fluids in large-scale structure simulations}},
  \href{https://doi.org/10.1088/1475-7516/2021/03/065}{\emph{JCAP} {\bfseries
  03} (2021) 065} [\href{https://arxiv.org/abs/2011.12503}{{\ttfamily
  2011.12503}}].

\bibitem{Chen:2022dsv}
J.~Z. Chen, M.~R. Mosbech, A.~Upadhye and Y.~Y.~Y. Wong, \emph{{Hybrid
  multi-fluid-particle simulations of the cosmic neutrino background}},
  \href{https://doi.org/10.1088/1475-7516/2023/03/012}{\emph{JCAP} {\bfseries
  03} (2023) 012} [\href{https://arxiv.org/abs/2210.16012}{{\ttfamily
  2210.16012}}].

\bibitem{Bayer:2020tko}
A.~E. Bayer, A.~Banerjee and Y.~Feng, \emph{{A fast particle-mesh simulation of
  non-linear cosmological structure formation with massive neutrinos}},
  \href{https://doi.org/10.1088/1475-7516/2021/01/016}{\emph{JCAP} {\bfseries
  01} (2021) 016} [\href{https://arxiv.org/abs/2007.13394}{{\ttfamily
  2007.13394}}].

\bibitem{Bond:1980ha}
J.~R. Bond, G.~Efstathiou and J.~Silk, \emph{{Massive Neutrinos and the Large
  Scale Structure of the Universe}},
  \href{https://doi.org/10.1103/PhysRevLett.45.1980}{\emph{Phys. Rev. Lett.}
  {\bfseries 45} (1980) 1980}.

\bibitem{Hu:1997vi}
W.~Hu and D.~J. Eisenstein, \emph{{Small scale perturbations in a general MDM
  cosmology}}, \href{https://doi.org/10.1086/305585}{\emph{Astrophys. J.}
  {\bfseries 498} (1998) 497}
  [\href{https://arxiv.org/abs/astro-ph/9710216}{{\ttfamily
  astro-ph/9710216}}].

\bibitem{Chevallier:2000qy}
M.~Chevallier and D.~Polarski, \emph{{Accelerating universes with scaling dark
  matter}}, \href{https://doi.org/10.1142/S0218271801000822}{\emph{Int. J. Mod.
  Phys. D} {\bfseries 10} (2001) 213}
  [\href{https://arxiv.org/abs/gr-qc/0009008}{{\ttfamily gr-qc/0009008}}].

\bibitem{Linder:2002et}
E.~V. Linder, \emph{{Exploring the expansion history of the universe}},
  \href{https://doi.org/10.1103/PhysRevLett.90.091301}{\emph{Phys. Rev. Lett.}
  {\bfseries 90} (2003) 091301}
  [\href{https://arxiv.org/abs/astro-ph/0208512}{{\ttfamily
  astro-ph/0208512}}].

\bibitem{Giare:2025ath}
W.~Giar{\`e}, O.~Mena, E.~Specogna and E.~Di~Valentino, \emph{{Neutrino mass
  tension or suppressed growth rate of matter perturbations?}},
  \href{https://doi.org/10.1103/njfc-pd1w}{\emph{Phys. Rev. D} {\bfseries 112}
  (2025) 103520} [\href{https://arxiv.org/abs/2507.01848}{{\ttfamily
  2507.01848}}].

\bibitem{Yang:2026qmz}
W.~Yang, S.~Zhang, D.~Barbosa, M.~A. Alvarez, S.~Pan and L.~Graef, \emph{{Do
  DESI-DR2 BAO data imply a coupling of dark matter and dark energy?}},
  \href{https://doi.org/10.1103/6kdf-vzq4}{\emph{Phys. Rev. D} {\bfseries 114}
  (2026) 063549} [\href{https://arxiv.org/abs/2609.05660}{{\ttfamily
  2609.05660}}].

\bibitem{Ambelu:2026mkd}
Y.~W. Ambelu, A.~Abebe, S.~B. Tessema and S.~Sahlu, \emph{{Alleviating the
  Hubble Tension Using Coupled Dark Energy - Dark Matter Interaction}},
  \href{https://arxiv.org/abs/2605.01904}{{\ttfamily 2605.01904}}.

\bibitem{Sabogal:2026bsw}
M.~A. Sabogal, S.~Vagnozzi and E.~Di~Valentino, \emph{{Cosmological anatomy of
  interacting dark energy}},
  \href{https://arxiv.org/abs/2609.18288}{{\ttfamily 2609.18288}}.

\bibitem{Stolzner:2025htz}
B.~St{\"o}lzner et~al., \emph{{KiDS-Legacy: Consistency of cosmic shear
  measurements and joint cosmological constraints with external probes}},
  \href{https://doi.org/10.1051/0004-6361/202554893}{\emph{Astron. Astrophys.}
  {\bfseries 702} (2025) A169}
  [\href{https://arxiv.org/abs/2503.19442}{{\ttfamily 2503.19442}}].

\bibitem{Pietroni:2008jx}
M.~Pietroni, \emph{{Flowing with Time: a New Approach to Nonlinear Cosmological
  Perturbations}},
  \href{https://doi.org/10.1088/1475-7516/2008/10/036}{\emph{JCAP} {\bfseries
  10} (2008) 036} [\href{https://arxiv.org/abs/0806.0971}{{\ttfamily
  0806.0971}}].

\bibitem{Lesgourgues:2009am}
J.~Lesgourgues, S.~Matarrese, M.~Pietroni and A.~Riotto, \emph{{Non-linear
  Power Spectrum including Massive Neutrinos: the Time-RG Flow Approach}},
  \href{https://doi.org/10.1088/1475-7516/2009/06/017}{\emph{JCAP} {\bfseries
  06} (2009) 017} [\href{https://arxiv.org/abs/0901.4550}{{\ttfamily
  0901.4550}}].

\bibitem{Upadhye:2017hdl}
A.~Upadhye, \emph{{Neutrino mass and dark energy constraints from
  redshift-space distortions}},
  \href{https://doi.org/10.1088/1475-7516/2019/05/041}{\emph{JCAP} {\bfseries
  05} (2019) 041} [\href{https://arxiv.org/abs/1707.09354}{{\ttfamily
  1707.09354}}].

\bibitem{Moran:2022iwe}
K.~R. Moran, K.~Heitmann, E.~Lawrence, S.~Habib, D.~Bingham, A.~Upadhye,
  J.~Kwan, D.~Higdon and R.~Payne, \emph{{The Mira{\textendash}Titan Universe
  {\textendash} IV. High-precision power spectrum emulation}},
  \href{https://doi.org/10.1093/mnras/stac3452}{\emph{Mon. Not. Roy. Astron.
  Soc.} {\bfseries 520} (2023) 3443}
  [\href{https://arxiv.org/abs/2207.12345}{{\ttfamily 2207.12345}}].

\bibitem{Upadhye:2023bgx}
A.~Upadhye, J.~Kwan, I.~G. McCarthy, J.~Salcido, K.~R. Moran, E.~Lawrence and
  Y.~Y.~Y. Wong, \emph{{Cosmic-E{\ensuremath{\nu}}: An- emulator for the
  non-linear neutrino power spectrum}},
  \href{https://doi.org/10.1093/mnras/stae783}{\emph{Mon. Not. Roy. Astron.
  Soc.} {\bfseries 530} (2024) 743}
  [\href{https://arxiv.org/abs/2311.11240}{{\ttfamily 2311.11240}}].

\end{thebibliography}\endgroup
\end{document}